\documentclass[aps,prl,twocolumn,superscriptaddress]{revtex4-2}
\pdfoutput=1
\usepackage{graphicx}
\usepackage{amsmath}
\usepackage{amssymb}
\usepackage{siunitx}
\usepackage[version=4]{mhchem}
\usepackage{lmodern}
\usepackage{gensymb}
\usepackage{xcolor}
\usepackage{color}
\usepackage{ragged2e}

\usepackage{booktabs}
\usepackage{etoolbox}
\usepackage{bibunits}
\usepackage[hidelinks,bookmarks=false]{hyperref}

\makeatletter
\auto@bib@empty
\newcommand{\printsupplementbibliography}{%
  \@startbibunitorrelax
  \if@filesw
    \immediate\write\@bibunitaux{\string\citation{REVTEX42Control,apsrev42Control}}%
    \immediate\write\@bibunitaux{\string\bibdata{\pre@bibdata,refs}}%
  \fi
  \let\auto@bib@innerbib\@empty
}

\newcommand{\suppressmaintoc}{%
  \let\original@addtocontents\addtocontents
  \renewcommand{\addtocontents}[2]{%
    \ifstrequal{##1}{toc}{}{\original@addtocontents{##1}{##2}}%
  }%
}

\newcommand{\startsupplement}{%
  \let\@hangfrom@section\@hang@from
  \let\@hangfroms@section\@hang@froms
  \setcounter{secnumdepth}{3}%
  \setcounter{section}{0}%
  \setcounter{subsection}{0}%
  \setcounter{subsubsection}{0}%
  \setcounter{figure}{0}%
  \setcounter{table}{0}%
  \setcounter{equation}{0}%
  \renewcommand{\thefigure}{S\arabic{figure}}%
  \renewcommand{\theHsection}{SI.\arabic{section}}%
  \renewcommand{\theHfigure}{SI.\arabic{figure}}%
  \renewcommand{\theHtable}{SI.\arabic{table}}%
  \renewcommand{\theHequation}{SI.\arabic{equation}}%
  \def\@extra@binfo{@SI}%
  \def\@extra@b@citeb{@SI}%
  \patchcmd{\endthebibliography}{\label{LastBibItem}}{\label{LastBibItemSI}}{}
    {\PackageError{merged-manuscript}{Unable to set the REVTeX bibliography label}{}}%
  \patchcmd{\NAT@bibsetnum}{\ref{LastBibItem}}{\ref*{LastBibItemSI}}{}
    {\PackageError{merged-manuscript}{Unable to set the REVTeX bibliography width}{}}%
  \let\@FMN@list\@empty
}
\makeatother

\begin{document}
\begingroup
\suppressmaintoc

\righthyphenmin=4
\lefthyphenmin=4

\title{Observation of nonadiabatic quantum tunneling in a driven correlated insulator}

\author{Mingyao Guo}
\affiliation{
 Department of Physics, California Institute of Technology, Pasadena, CA 91125, USA
}
\affiliation{
 Institute for Quantum Information and Matter, California Institute of Technology, Pasadena, CA 91125, USA
}

\author{Radu Andrei}
\affiliation{
 Institute for Theoretical Physics, ETH Zurich, 8093 Zurich, Switzerland
}

\author{Jaena Park}
\affiliation{
 Center for Quantum Materials, Department of Physics and Astronomy, Seoul National University, Seoul 08826, Korea
}

\author{Je-Geun Park}
\affiliation{
 Center for Quantum Materials, Department of Physics and Astronomy, Seoul National University, Seoul 08826, Korea
}
 
\author{Eugene Demler}
\affiliation{
 Institute for Theoretical Physics, ETH Zurich, 8093 Zurich, Switzerland
}
 
\author{David Hsieh}
\email{dhsieh@caltech.edu}
\affiliation{
 Department of Physics, California Institute of Technology, Pasadena, CA 91125, USA
}
\affiliation{
 Institute for Quantum Information and Matter, California Institute of Technology, Pasadena, CA 91125, USA
}

\begin{abstract}

Many emerging solid-state platforms for Floquet engineering, nonlinear transport, and high-harmonic generation operate near the Keldysh crossover, where optical fields are sufficiently strong and fast to reshape electronic states on sub-cycle timescales. Despite its importance, the microscopic pathways for photocarrier generation in this regime remain poorly understood, as neither perturbative multiphoton absorption nor adiabatic tunneling provides an adequate description. Here, we perform time-resolved optical spectroscopy on the correlated van der Waals insulator NiPS$_3$ while systematically tuning the drive electric field and frequency over a wide range. In the crossover regime, we observe multiphoton-like resonances that are blue-shifted by ponderomotive effects within an otherwise tunneling-dominated absorption channel — a clear spectroscopic signature of adiabaticity breakdown. Hubbard-model simulations based on a time-dependent Gaussian variational approach qualitatively reproduce these features. Together, our results elucidate the microscopic mechanisms of charge excitation in the Keldysh crossover regime and identify a key ingredient for controlling strong-field nonlinear dynamics in correlated materials.

\end{abstract}

\maketitle
\date{\today}

\section{Introduction}
Strong off-resonant optical driving of complex insulating quantum materials offers a powerful route to controlling electronic properties far from equilibrium \cite{basov2017towards,de2021colloquium}, with implications for lightwave-based electronics and spintronics \cite{borsch2023lightwave,heide2024petahertz}. A central problem is understanding photocarrier generation in the sub-gap driving regime, with an early theoretical description formulated by Keldysh \cite{keldysh1964ionization} and subsequently extended in related works \cite{popov2004tunnel,kruchinin2018colloquium}. In this solid-state analogue of atomic ionization, the binding potential is replaced by the bandgap, and the strong-field drive creates an electron-hole pair rather than a free electron and a residual ion. In this framework, charge excitations are controlled by a dimensionless adiabaticity (Keldysh) parameter $\gamma_K$, defined as the ratio of the Keldysh tunneling time — the time for a semi-classical electron to traverse a static barrier — to the laser period. Up to a material-dependent prefactor, this yields $\gamma_K\propto\omega/E$, where $\omega$ and $E$ are the light field frequency and amplitude, respectively. In the limit where $\gamma_K \ll 1$, electrons experience the energy barrier as quasi-static and therefore tunnel adiabatically (Fig.~\ref{fig:m1}a, left inset). On the other hand, when $\gamma_K \gg 1$, photo-excitation proceeds via a multiphoton process and thus exhibits a strong frequency dependence (Fig.~\ref{fig:m1}a, right inset). The intermediate regime $\gamma_K \sim 1$, dubbed the ``Keldysh crossover'', underpins a variety of sub-gap light-induced phenomena, such as Floquet engineering \cite{shan2021giant,zhou2023pseudospin} and high-harmonic generation \cite{ghimire2011observation,heide2024ultrafast}. Yet it lies beyond standard multiphoton and adiabatic tunneling frameworks, and its microscopic origin remains largely unexplored.

Although signatures of the Keldysh crossover have been reported in Ca$_2$RuO$_4$ \cite{li2022keldysh} and  CsPbBr$_3$ \cite{zhang2025keldysh}, these studies primarily identified tunneling-like or multiphoton-like scaling behaviors in separate regions of parameter space. As a result, the microscopic charge excitation dynamics in the crossover regime — where tunneling and multiphoton channels cooperate within a single optical cycle — remain almost uncharted. Here, the intense AC field transiently tilts the potential barrier but reverses direction on a timescale comparable to the tunneling time, rendering the excitation process intrinsically nonadiabatic \cite{yudin2001nonadiabatic,ivanov2005anatomy}. In principle, attosecond time-domain techniques such as attoclock measurements based on angular streaking \cite{eckle2008attosecond1,eckle2008attosecond2,sainadh2019attosecond} and tunneling interferometry \cite{shafir2012resolving,pedatzur2015attosecond,kneller2022look} can access such sub-cycle dynamics, but their reliance on free-electron propagation and detection of very high-order harmonics has so far limited their applicability to atomic and molecular systems. 

An alternative route to interrogating nonadiabatic tunneling dynamics in solids lies in the frequency domain. Canonical Keldysh theory predicts that near $\gamma_K \sim 1$, the tunneling excitation rate acquires a pronounced dependence on photon energy, giving rise to resonant enhancements that resemble multiphoton absorption despite originating from a nonperturbative tunneling process \cite{keldysh1964ionization}. Systematically tracking the photocarrier generation rate as a function of drive frequency in this intermediate regime therefore provides a powerful spectroscopic handle on nonadiabatic tunneling. However, implementing this approach experimentally requires precise and highly sensitive measurements, as sub-gap tunneling generates relatively modest photocarrier densities whose frequency dependence must be resolved quantitatively over a wide range of field strengths.

In this Letter, we realize this frequency-domain approach by performing transient reflectivity measurements on the antiferromagnetic insulator NiPS$_3$, whose strong correlations and van der Waals structure — endowing the crystal with a soft out-of-plane elastic modulus and reduced dielectric screening — enhance both photocarrier generation and coherent longitudinal acoustic phonons that sensitively probe the photo-excited electron-hole pair density $n$. By systematically tracking $n$ across the $\gamma_K \sim 1$ regime, we observe pronounced resonant features at frequencies near integer fractions of the bandgap that resemble multiphoton absorption, while their field dependence reveals a tunneling-dominated excitation mechanism. These resonances exhibit substantial blue-shifts arising from the ponderomotive energy associated with strong-field dressing of the photocarriers. Hubbard model simulations employing a time-dependent Gaussian variational approach qualitatively capture these trends. Collectively, the observed phenomenology bridges a gap between conventional nonlinear absorption regimes in gapped systems and offers new insights into coherent light control of quantum materials with minimal dissipation.

\section{Results}
The differential reflectivity ($\Delta R/R$) of bulk single crystals of NiPS$_3$ was measured as a function of time delay (\emph{t}) between an ultrashort pump pulse ($\sim$100 fs) with tunable photon energy ($\hbar\omega$) and a weak probe pulse centered at 1.55 eV. The charge-transfer gap of NiPS$_3$ is $\Delta = 1.8$ eV \cite{kim2018charge} (Fig.~\ref{fig:m1}a); we therefore begin by pumping well above the gap at 2 eV, where photocarriers are generated via simple linear absorption, allowing us to calibrate the dependence of our observables on $n$. Following excitation, we observe a rapid drop of $\Delta R/R$ (Fig.~\ref{fig:m1}b). The maximal magnitude of $\Delta R/R$ is proportional to $n$ \cite{demsar1999single,chia2006quasiparticle,li2022keldysh} and scales quadratically with the peak electric field $E$ (Fig.~\ref{fig:m1}c), as expected for linear optical absorption. 

\begin{figure}[!htb]
\includegraphics{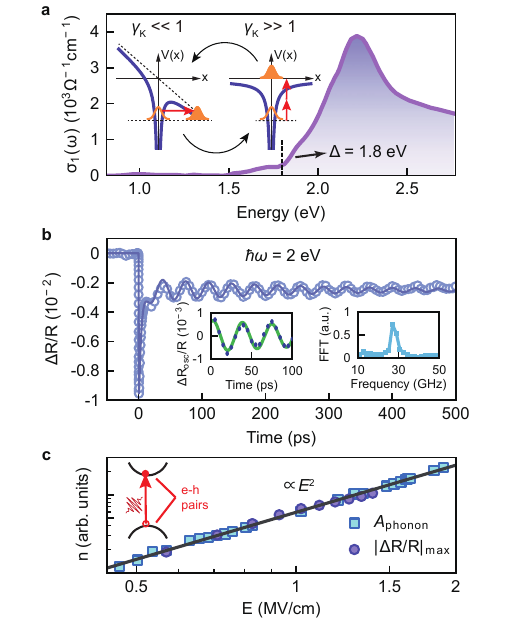}
\caption{\label{fig:m1} Photocarrier generation in NiPS$_3$ by linear absorption. (a) Real part of the optical conductivity $\sigma_1(\omega)$ of NiPS$_3$ at 100 K. The inset shows schematic real-space representations of the Coulomb potential barrier in the adiabatic tunneling and multiphoton regimes. (b) 2 eV pump $\Delta R/R$ trace at $E = 1.9$ MV/cm. The solid line is a fit to the sum of a bi-exponential decay, a damped cosine function, and a constant. A background-subtracted transient  (green curve shows fit to damped cosine function) and its Fourier spectrum are shown in the insets. (c) Absolute value of reflectivity transient extremum $|\Delta R/R|_{\rm{max}}$ near time-zero (circles) and phonon amplitude $A_{\rm{phonon}}$ (squares) as a function of $E$ at $\hbar\omega = 2$ eV on a log-log scale. The solid line is the expected scaling for a linear absorption process as depicted in the inset.}
\end{figure}

\begin{figure*}[!htb]
\includegraphics{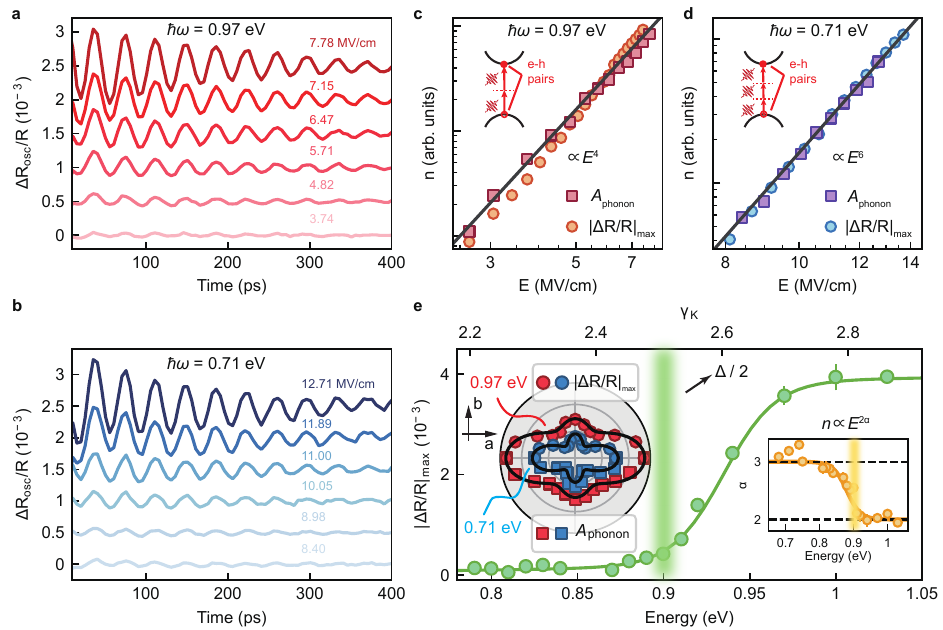}
\caption{\label{fig:m2} Photocarrier generation in NiPS$_3$ by multiphoton absorption. (a) Background-subtracted reflectivity transients at various peak electric fields for $\hbar\omega$ = 0.97 eV and (b) 0.71 eV. Curves are vertically offset for clarity. (c) $n$ versus $E$ for $\hbar\omega$ = 0.97 eV and (d) 0.71 eV on a log-log scale. Data points extracted from $|\Delta R/R|_{\rm{max}}$ and $A_{\rm{phonon}}$ are labeled by circles and squares, respectively. The solid lines show the expected scaling behaviors of the corresponding two- and three-photon processes as depicted in the insets. (e) $|\Delta R/R|_{\rm{max}}$ versus $\hbar\omega$ (bottom axis) and $\gamma_K$ (top axis) across $\Delta / 2$ for $E = 7.5$ MV/cm. The green curve is a guide to the eye. The left inset shows a polar plot of $n$ as a function of linear pump polarization angle relative to the crystal axes with $\hbar\omega$ = 0.97 eV (red symbols) and 0.71 eV (blue symbols) determined by $|\Delta R/R|_{\rm{max}}$ (circles, upper half) and $A_{\rm{phonon}}$ (squares, bottom half). Black lines are fits to the $2/m$ crystal symmetry. The right inset shows the $\hbar\omega$ dependence of the multiphoton exponent $\alpha$ extracted by fitting to $n\propto E^{2\alpha}$ across $\Delta / 2$. The applicability of the Keldysh theory in this low-order multiphoton regime is discussed in Supplementary Text S\ref{sec:Keldysh theory}.}
\end{figure*}

Superposed on the recovery of $\Delta R/R$ are coherent oscillations at 27 GHz frequency that persist over nanosecond timescales (Fig.~\ref{fig:m1}b, right inset). After subtracting the incoherent electronic background (Supplementary Text S\ref{sec:fit trr}), the oscillatory component is well described by a damped cosine (Fig.~\ref{fig:m1}b, left inset), indicative of displacive excitation. The observed frequency agrees with recent assignments of this mode to a longitudinal acoustic phonon \cite{wright2024acoustic,sahu2025ultrafast}. On these intermediate timescales, photocarriers relax their excess energy into the phonon bath \cite{thomsen1984coherent,thomsen1986surface}, rendering the coherent phonon amplitude proportional to the total absorbed energy. Accordingly, the oscillation amplitude also exhibits a quadratic dependence on $E$ (Fig.~\ref{fig:m1}c), corroborating the scaling of $\Delta R/R$ at fixed delay. We therefore identify both the transient reflectivity amplitude and the coherent phonon response as independent relative probes of $n$, and use them interchangeably throughout this work. Their microscopic transduction mechanisms and the robustness of their field and photon-energy dependences are discussed in Supplementary Text S\ref{sec:probe}.

\begin{figure*}[!htb]
\includegraphics{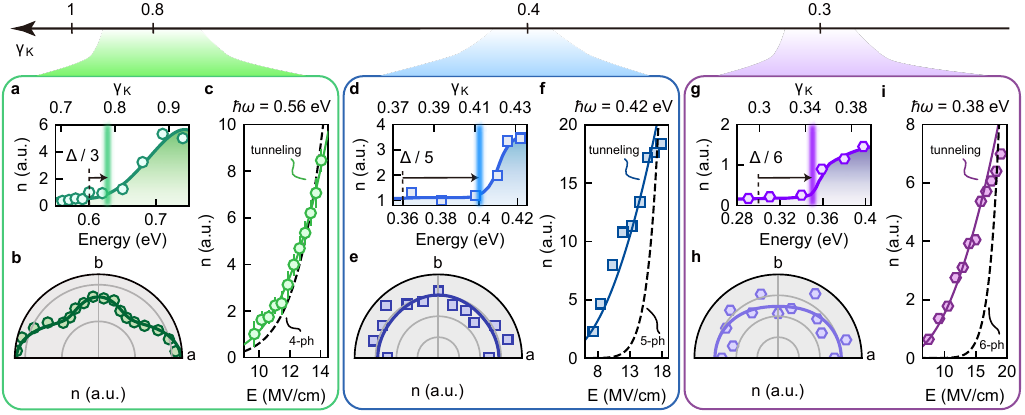}
\caption{\label{fig:m3} Photocarrier generation in NiPS$_3$ by quantum tunneling. (a) $n$ as a function of $\hbar\omega$ (bottom axis) and $\gamma_K$ (top axis) across $\Delta/N$ for $N$ = 3, (d) $N$ = 5, (g) $N$ = 6 at $E=14, 17.8$, and 19 MV/cm, respectively. Blueshifts are labeled by the arrows. Solid lines are guides to the eye. (b) Polar plots of $n$ at various linear pump polarization angles with $\hbar\omega$ = 0.56 eV, (e) 0.42 eV, and (h) 0.38 eV and the same $E$ as in (a), (d), and (g). Solid lines are fits based on the crystal symmetry. (c) $n$ versus $E$ with $\hbar\omega$ = 0.56 eV, (f) 0.42 eV and (i) 0.38 eV. Solid lines are fits to the tunneling formula, while dashed lines correspond to power-law fits with the expected multiphoton order (Supplementary Text S\ref{sec:datasets}). In (a)-(c), $n$ is determined from $|\Delta R/R|_{\rm{max}}$ and in (d)-(i) it is determined from $A_{\rm{phonon}}$. The top axis schematically indicates the magnitude of $\gamma_K$ for each set of measurements (Supplementary Text S\ref{sec:Keldysh parameter}).}
\end{figure*}

We first examined sub-gap photocarrier generation in the multiphoton limit ($\gamma_K \gg 1$) using a drive photon energy $\Delta/2<\hbar\omega = 0.97$ eV $<\Delta$. Excitation at this frequency produces an abrupt drop in reflectivity (fig.~\ref{fig:f dep}) accompanied by coherent phonon oscillations (Fig.~\ref{fig:m2}a), qualitatively similar to the above-gap response but requiring substantially higher fluences. The amplitudes of both the reflectivity change and the phonon oscillations follow a clear power-law scaling $n\propto E^{2\alpha}$ with $\alpha\sim 2$ (Figs.~\ref{fig:m2}c and~\ref{fig:field dependence 2 ph 3 ph}), consistent with a two-photon process expected from this photon energy. 

At a lower drive frequency $\Delta/3<\hbar\omega = 0.71$ eV $<\Delta/2$, an analogous response is observed (Fig.~\ref{fig:m2}b); however, $n$ now exhibits a cubic intensity dependence with $\alpha\sim 3$ (Figs.~\ref{fig:m2}d and~\ref{fig:field dependence 2 ph 3 ph}), indicating the dominance of three-photon excitations. Scanning the pump frequency across $\Delta/2$ at a fixed peak electric field $E$ reveals a sharp crossover between these regimes, manifested as a step-like change in $n$ (Fig.~\ref{fig:m2}e) and a concomitant evolution of the power-law exponent from $\alpha$ = 2 to $\alpha$ = 3 at the threshold energy (Fig.~\ref{fig:m2}e, right inset). 

As a hallmark of perturbative optics, multiphoton absorption is governed by high-order susceptibility tensors that reflect the symmetries of the crystallographic point group \cite{hurlbut2007multiphoton, pearl2008three, gertsvolf2008orientation, kozak2019anisotropy}. Consistent with this expectation, rotating the in-plane pump polarization produces a pronounced anisotropy in $n$ at both 0.97 eV and 0.71 eV, in agreement with the $2/m$ symmetry of NiPS$_3$ \cite{ouvrard1985structural,brec1986review} (Fig.~\ref{fig:m2}e, left inset).

Approaching the Keldysh crossover, the photocarrier response is expected to develop three distinct signatures. First, tunneling characteristics begin to emerge. In two dimensions, interband tunneling in the quasi-static limit ($\gamma_K\ll 1$) exhibits a threshold behavior $n\propto E^2\exp{(-E_{\rm{th}}/E)}$. Second, as the drive period becomes comparable to the Keldysh tunneling time ($\gamma_K \sim 1$), the adiabatic approximation breaks down, and the drive frequency is anticipated to strongly influence the tunneling-induced photocarrier density. Third, in this non-perturbative tunneling regime, the symmetry of charge excitations is expected to evolve: selection rules imposed by the crystallographic point group progressively weaken, giving way to an increasingly isotropic response as the strong optical field tilts the real-space Coulomb potential sufficiently that the pair production rate becomes insensitive to crystal orientation (Supplementary Text S\ref{sec:datasets}).

We experimentally access the Keldysh crossover regime in NiPS$_3$ by further detuning the pump to lower energies while simultaneously increasing the field strength. Figure~\ref{fig:m3} summarizes the resulting dependence of $n$ on excitation energy and field strength in three representative regions of the crossover. As shown in Fig.~\ref{fig:m3}a, sweeping the pump energy across the nominal three- to four-photon threshold near $\Delta/3$ produces a pronounced step in $n$, reminiscent of the transition observed at $\Delta/2$. However, the estimated Keldysh parameter satisfies $\gamma_K < 1$ throughout this range, suggesting that photocarrier generation is dominated by tunneling. The field dependence of $n$ at $\hbar\omega = 0.56$ eV, which lies in the range  $\Delta/4<\hbar\omega<\Delta/3$, deviates from the power-law scaling expected for four-photon absorption and is instead quantitatively captured by the tunneling form introduced above (Fig.~\ref{fig:m3}c), signaling a crossover from a three-photon regime to a tunneling-dominated regime as the drive frequency decreases. Polarization-resolved measurements further corroborate this picture. Whereas $n$ exhibits pronounced anisotropy in the multiphoton regime (Fig.~\ref{fig:m2}e, left inset), this anisotropy is markedly suppressed at $\hbar\omega = 0.56$ eV (Fig.~\ref{fig:m3}b): the carrier response becomes comparable along the $a$ and $b$ axes, consistent with the increasingly isotropic character expected for tunneling-driven excitation.

\begin{figure*}[!htb]
\includegraphics{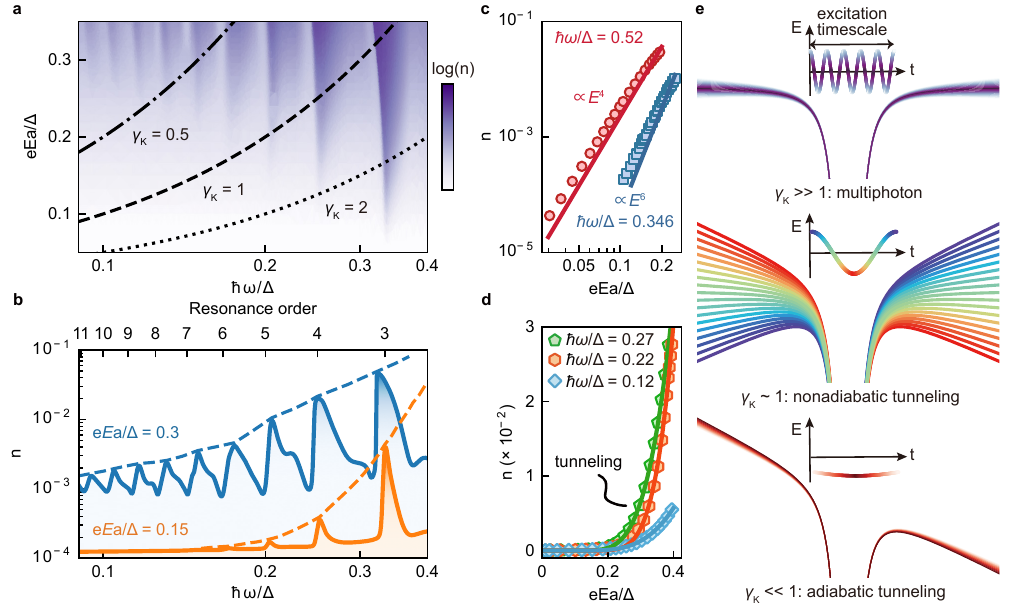}
\caption{\label{fig:m4} Driven Hubbard model numerics. (a) Simulated $n$ versus drive energy $\hbar\omega/\Delta$ and field amplitude $eEa/\Delta$ on a log scale. Curves of constant $\gamma_K$ are overlaid. (b) Select constant $E$ cuts through the data in (a) at $eEa/\Delta = 0.15, 0.3$. Over the displayed interval, these cuts span approximately $0.6\lesssim\gamma_K\lesssim2.7$ and $0.3\lesssim\gamma_K\lesssim1.3$, respectively. Dashed lines are guides to the eye. (c) Select constant $\hbar\omega$ cuts at $\hbar\omega/\Delta = 0.346, 0.52$ overlaid with their expected power-law scaling on a log-log scale. (d) Select constant $\hbar\omega$ cuts at $\hbar\omega/\Delta = 0.27, 0.22,$  and 0.12 overlaid with fits to the tunneling formula. (e) Schematic real-space representation of the field-driven distortion of the potential barrier across the multiphoton ($\gamma_K\gg 1$, top), nonadiabatic tunneling ($\gamma_K\sim1$, middle), and adiabatic tunneling ($\gamma_K\ll 1$, bottom) regimes. The electric field (insets) is shown up to the charge excitation timescale and the barrier is sketched at select times using colors that correspond to the field traces.}
\end{figure*}

Reducing the pump frequency to still lower photon energies produces analogous step-like features in $n$ near $\Delta/5$ and $\Delta/6$, driving the system further into the tunneling regime (Fig.~\ref{fig:m3}d, g). In these regimes, the measured field dependence of $n$ quantitatively follows the tunneling form, again confirming its non-perturbative origin (Fig.~\ref{fig:m3}f,i). Concurrently, $n$ becomes increasingly insensitive to the pump polarization (Fig.~\ref{fig:m3}e,h), providing further evidence for the dominance of tunneling-driven excitation. Impact ionization and avalanche multiplication can in principle follow primary strong-field injection; however, quantitative estimates of carrier energy, threshold field, and pulse intensity indicate that they make at most a minor contribution under our driving conditions (Supplementary Text S\ref{sec:impact ionization}).

The pronounced frequency dependence of the tunneling yield observed here extends beyond the textbook picture of adiabatic tunneling and constitutes direct experimental evidence of nonadiabatic quantum tunneling. This behavior is in excellent agreement with Keldysh theory, which predicts a strong sensitivity of the excitation rate to photon energy for $\gamma_K \sim 1$. Such dynamics are unique to electrons driven by a strong AC field and continuously interpolate between the perturbative multiphoton regime at higher frequencies and weaker fields, and the adiabatic tunneling regime at lower frequencies and stronger fields.

Notably, all the multiphoton-like resonances observed in the $\gamma_K< 1$ regime are systematically blue-shifted relative to the exact integer fractions of the equilibrium bandgap (Fig.~\ref{fig:m3}a, d, g). This shift arises from the ponderomotive energy $E_p$ acquired by electrons and holes in a strong electromagnetic field, which renormalizes the effective gap to $\tilde{\Delta} = \Delta + E_p$ \cite{kruchinin2018colloquium}. Using the experimental field amplitudes and a reduced effective mass $m^*\approx0.45m_e$ obtained from density functional theory \cite{lane2020thickness}, we estimate $E_p=e^2E^2/4m^*\omega^2$ to be approximately 54 meV, 210 meV, and 312 meV under the excitation conditions of Fig.~\ref{fig:m3}a, d, g, respectively. These values predict renormalized resonance energies at 0.62 eV ($\tilde{\Delta}/3$), 0.4 eV ($\tilde{\Delta}/5$), and 0.35 eV ($\tilde{\Delta}/6$), in quantitative agreement with the experimentally observed thresholds. Detailed resonance analysis and temperature-dependent measurements across $T_N$ supporting this assignment are presented in Supplementary Text S\ref{sec:t dep}.


To support our findings, we numerically simulated the photocarrier density $n$ using drive and materials parameters comparable to those in our experiments. We employed a time-dependent Gaussian variational approach to compute the charge dynamics of a two-dimensional repulsive Hubbard model at half filling with $U/\tau=10$ following ultrafast subgap excitation (see Materials and Methods for details) \cite{andrei2025subgap}. Although minimal, this model captures the essential local electron-electron interactions relevant to NiPS$_3$ and treats the light-matter interaction in a non-perturbative manner, while enabling both self-consistent numerical simulations and analytic approximations that reproduce the key features of the excitation dynamics across the Keldysh crossover.

Figure~\ref{fig:m4}a presents a comprehensive map of calculated $n$ as a function of photon energy and peak electric field on a logarithmic scale. In the $\gamma_K>1$ regime, multiphoton absorption manifests as discrete peaks in $n$ at energies near integer fractions of the gap $\Delta$, whose amplitudes rapidly diminish for increasing orders due to the perturbative nature of the process (Fig.~\ref{fig:m4}b, orange curve). Their corresponding field dependences follow power-law scaling with the expected exponents (Fig.~\ref{fig:m4}c), confirming their multiphoton character. Crucially, the simulations reveal that as $\gamma_K$ decreases toward and below unity, these resonance features persist deep into the quantum tunneling regime. Even for $\gamma_K<1$, pronounced step-like structures remain visible near nominal multiphoton resonances. However, their amplitudes no longer decay exponentially with increasing order and instead approach a near-plateau at low frequencies (Fig.~\ref{fig:m4}b, blue curve), highlighting their nonperturbative origin.

The corresponding field dependences likewise follow the tunneling scaling law (Fig.~\ref{fig:m4}d), confirming the dominance of tunneling-driven charge excitation in this regime. We note that nearest-neighbor square-lattice dispersion produces a sharp density of states at the band edge (fig.~\ref{fig:density_of_states}); in NiPS$_3$, where the absorption spectrum is considerably smoother (Fig.~\ref{fig:m1}), the same features evolve into step-like responses in experiments. Taken together, these numerical results closely mirror our experimental findings and establish that photocarrier generation near the Keldysh crossover is governed by nonadiabatic tunneling, as reflected simultaneously in tunneling-type field dependences and a pronounced photon-energy sensitivity. These behaviors are also qualitatively consistent with prior numerical studies of driven Hubbard models based on Floquet dynamical mean-field theory \cite{murakami2018nonequilibrium,murakami2025photoinduced}. The slight redshifted resonance edges at high fields in the simulations arise from self-consistent photocarrier-induced suppression of the correlation gap (Supplementary Text S\ref{sec:Hubbard}). This effect plays only a minor role in our experiments due to the relatively weak dielectric screening in van der Waals NiPS$_3$, and is discussed elsewhere \cite{andrei2025subgap}. In addition, we performed complementary toy-model calculations within the Keldysh framework (Supplementary Text S\ref{sec:Keldysh theory}) and reproduced both signatures of nonadiabatic tunneling and the associated ponderomotive blueshift (fig.~\ref{fig:Keldysh theory}), further reinforcing our interpretation.

\section{Discussion}
This work completes the picture of how sub-gap lightwaves with varying photon energies and field strengths generate charge excitations in an insulating material. In the high frequency, low field limit ($\gamma_K \gg 1$), electrons experience a rapidly modulated but weakly distorted potential and absorb multiple photons until they acquire sufficient energy to overcome the gap (Fig.~\ref{fig:m4}e, top panel). By contrast, at moderate frequencies and stronger fields where the potential is distorted on atomic length scales ($\gamma_K \sim 1$), electrons tunnel through a slowly oscillating barrier in a nonadiabatic manner (Fig.~\ref{fig:m4}e, middle panel). In this regime, the photocarrier density exhibits resonant enhancements when the drive photon energy exceeds an integer fraction of the ponderomotive-energy-renormalized gap $\tilde{\Delta}$. Our combined experimental and theoretical results demonstrate that this nonadiabatic tunneling regime remains robust down to at least $\gamma_K \sim 0.3$, a parameter range directly relevant to a wide class of strong-field phenomena in solids \cite{kruchinin2018colloquium,borsch2023lightwave,heide2024petahertz,heide2024ultrafast}. In the asymptotic quasi-static limit ($\gamma_K\ll 1$), tunneling occurs on a timescale much shorter than the field oscillation period, and the system evolves adiabatically (Fig.~\ref{fig:m4}e, bottom panel), recovering the frequency-independent excitation rate (fig.~\ref{fig:Keldysh theory}b).

There are numerous theoretical proposals for defining the electron tunneling time \cite{hauge1989tunneling,landauer1994barrier}, and the observation of a finite nonadiabatic timescale in our measurements suggests that the Keldysh time — formally equivalent to the Büttiker-Landauer traversal time \cite{buttiker1982traversal} — provides a physically relevant characterization of tunneling dynamics in solids \cite{mcdonald2013tunnel}. Beyond its conceptual significance, this insight has direct implications for other strong-field phenomena governed by quantum tunneling, most notably high-harmonic generation. Within the standard three-step picture, electrons acquire a finite initial velocity and accumulate a nontrivial quantum phase while traversing the barrier nonadiabatically, resulting in exit points that are shifted relative to the associated holes. These effects are expected to modify the sub-cycle timing of harmonic emission and leave direct fingerprints on the resulting spectrograms \cite{zuo2023revealing}. In parallel, harmonics generated during the tunneling process itself — so-called Brunel harmonics \cite{brunel1990harmonic,jurgens2020origin} — can serve as an all-optical attoclock for tracking the evolution of tunneling wavepackets \cite{babushkin2022all}. Future experiments employing time-domain high-harmonic spectroscopy \cite{luu2015extreme,hohenleutner2015real,langer2017symmetry,uchida2024high} and Brunel radiation may thus enable real-time observation of tunneling dynamics in solids, offering direct insight into the nonadiabatic tunneling process in the time domain.

\section{Materials and Methods}
\subsection{Crystal synthesis}
High-quality NiPS$_3$ single crystals were grown using the chemical vapor transport method, following our previous work \cite{kuo2016exfoliation}. Ni, P, S, and I$_2$ were mixed in an argon-filled glove box and sealed in a quartz tube under vacuum. The quartz tube was placed in a two-zone furnace with a temperature gradient of 750 $^\circ$C (source) to 700 $^\circ$C (sink) for 7 days. We verified the stoichiometry of the synthesized single crystals with a COXEM-EM30 scanning electron microscope equipped with a Bruker QUANTAX 70
energy-dispersive X-ray system. We also confirmed the quality of the samples by measuring Raman spectrum at room temperature (Nanobase XPER-RAM-35V). The crystal structure was confirmed through X-ray diffraction measurements using a commercial diffractometer (Rigaku Miniflex II).

The samples were cleaved along the
[001] direction before the optical experiment to obtain a fresh and smooth surface and were then kept in high vacuum with pressure lower than 7 $\times$ 10$^{-7}$ torr during transient reflectivity measurements.

\subsection{Time-resolved reflectivity measurements}
Time-resolved reflectivity experiments were performed using a Ti:Sapphire amplified laser system with 35 fs pulses at a central photon energy of 1.55 eV and a repetition rate of 1 kHz. Part of the fundamental of the laser was split off into a dual-stage optical parametric amplifier (OPA), which outputs pulses in the near-infrared with photon energies ranging from 0.56 to 1.03 eV. The signal and idler beams of the OPA were further sent into a difference frequency generator in the non-collinear geometry to generate pulses in the mid-infrared with photon energies ranging from 0.29 to 0.42 eV. Both near-infrared and mid-infrared pulses have $\sim$100 fs pulse duration at the sample position (Supplementary Text S\ref{sec:pulse duration}). The pump pulses were incident on the sample at an angle of $\sim$15$\degree$ away from the normal direction (details in Supplementary Text S\ref{sec:oblique incidence}) and the pump spot sizes were 130 $\mu$m in diameter. The probe pulses at 800 nm have a duration of 65 fs at the sample position. The probe beam was focused obliquely onto the sample at an angle of 10$\degree$ with a spot size of 80 $\mu$m in diameter and a fluence lower than 10 $\mu$J/cm$^2$. The pump and probe lasers are linearly polarized along $b$ and $a$ axes respectively unless mentioned otherwise (Supplementary Text S\ref{sec:orientation}). The fluences throughout the work are calculated based on using 1/$e^2$ width as the beam diameter. All experiments were performed at 90 K unless otherwise specified and no measurable change of phonon frequency was resolved across our entire fluence range.

\subsection{Hubbard model simulations}
We study nonlinear charge dynamics in a two-dimensional half-filled Hubbard model on a square lattice, by solving its time evolution under optical driving, via a time-dependent Gaussian variational approach. We use the Fermi-Hubbard Hamiltonian
\begin{equation}
    H(t) = -\tau\sum_{j, \hat{\delta}, \alpha}  e^{ie \hat{\delta} \cdot \vec{A}(t)} \; c_{j+\hat{\delta}, \alpha}^\dagger c_{j, \alpha} + U \sum_j n_{j,\uparrow} n_{j,\downarrow},
\end{equation}
where $\tau$ is the electron hopping amplitude between nearest neighbors, $j$ indexes lattice sites, $\hat{\delta} \in \{\pm x, \pm y\}$ denotes nearest-neighbor directions, $U$ is the on-site Coulomb repulsion, $\alpha$ is the electron spin, and the laser field vector potential $A(t)$ enters through the time-dependent Peierls phase. The model parameters are set to $U/\tau=10$ to mimic the correlation effects in NiPS$_3$. The laser pulse duration is taken as 100 fs to match the experimental conditions.

We employ a variational approach starting from the Mott insulating ground state, which is characterized by charge localization as well as a tendency towards AFM ordering with N\'{e}el wavevector $\bold{Q}=(\pi,\pi)$. At the Gaussian (self-consistent mean-field) level, the relevant time-dependent wavefunction is
\begin{equation} \label{eq:methods_variational_wavefunction}
    |\psi(t) \rangle = \prod_{\bold{k}\in\rm{BZ^\prime},\alpha} \Big( u_{\bold{k}} (t) \; c_{\bold{k}, \alpha}^\dagger + v_{\bold{k}}(t) \; c_{\bold{k+Q}, \alpha}^\dagger \Big) |0\rangle,
\end{equation}
where $|0\rangle$ is the fermionic vacuum state, and BZ$^\prime$ denotes the AFM Brillouin zone. The initial coefficients $u_{\bold{k}} (0), v_{\bold{k}}(0)$ are chosen to minimize the energy expectation, i.e. we start in the ground state of the variational manifold. For $t>0$, the system state evolves under the drive $A(t)$. Note that we extend the wavefunction \eqref{eq:methods_variational_wavefunction} to a density-matrix based approach, in order to also include decoherence effects present in solid-state systems.  We calculate the density of incoherent doublons and holons after the pump pulse ($t> T_{\rm pump}$), and use it in this work. For more details, see Supplementary Text S\ref{sec:Hubbard}.

\section{Acknowledgments}
We acknowledge helpful discussions with Xinwei Li, Carina Belvin, and Tsing-Hua Her. 

\section{Funding}
M.G. was supported by a California Institute of Technology Zewail Fellowship. D.H. acknowledges support from a Brown Investigator Award, a program of the Brown Institute for Basic Sciences at the California Institute of Technology, and the Institute for Quantum Information and Matter, an NSF Physics Frontiers Center (PHY-1733907). E.D. and R.A. acknowledge support from the SNSF project 200021\_212899, the Swiss State Secretariat for Education, Research and Innovation (contract number UeM019-1), and NCCR SPIN, a National Centre of Competence in Research, funded by the Swiss National Science Foundation (grant number 225153). J.P. and J.-G.P. were supported by the Leading Researcher Program of the National Research Foundation of Korea (Grant Nos. RS-2020-NR049405).

\section{Author contributions}
M.G. performed optical experiments and data analysis. R.A. carried out Hubbard model numerical simulations supervised by E.D. J.P. and J.-G.P. grew and characterized the single crystals. M.G. and D.H. wrote the manuscript with contribution from all other authors. D.H. supervised the project.

\section{Competing interests}
The authors declare no competing interests.

\section{Data, code, and materials availability}
All data and code needed to evaluate and reproduce the results in the paper are present in the paper and/or the Supplementary Materials. This study did not generate new materials.

\bibliography{refs}
\clearpage
\endgroup

\onecolumngrid
\begin{bibunit}[apsrev4-2]
\startsupplement

\begin{center}
  {\large\bfseries Supplementary Materials:\\Observation of nonadiabatic quantum tunneling in a driven correlated insulator\par}
  \medskip
  Mingyao Guo, Radu Andrei, Jaena Park, Je-Geun Park, Eugene Demler, and David Hsieh
\end{center}

\tableofcontents

\setlength{\parindent}{2em}
\setlength{\parskip}{0.5em}
\setlength{\abovedisplayskip}{1.75em}
\setlength{\belowdisplayskip}{1.75em}
\setlength{\intextsep}{20pt plus 1.0pt minus 2.0pt}
\renewcommand \thesection{\arabic{section}}

\clearpage
\section{Determination of crystal orientation}
\label{sec:orientation}
\begin{figure}[!htb]
\includegraphics[width=0.3\linewidth]{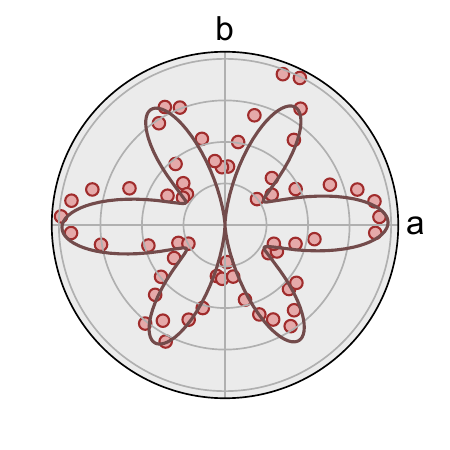}
\caption{\label{fig:shg} SHG pattern of NiPS$_3$ in the S$_{\rm{in}}$-S$_{\rm{out}}$ geometry at 250~K. The solid line is a fit to the data based on the bulk electric quadrupole and surface electric dipole radiations.}
\end{figure}

The orientation of bulk NiPS$_3$ single crystals is determined using rotational anisotropy second harmonic generation \cite{harter2015high} (SHG). An 800 nm laser beam is used as the light source and its second harmonic is collected in the reflection geometry. We adopted S$_{\rm{in}}$-S$_{\rm{out}}$ configuration and the resulting SHG pattern at 250~K above the N\'{e}el temperature is shown in Fig.~\ref{fig:shg}. Based on the bulk $2/m$ symmetry \cite{chu2020linear}, we determine the corresponding crystallographic axes.

When fitting the polarization dependence of $n$ in the sub-gap regime, we expanded it in terms of the crystal mirror symmetry perpendicular to the $b$ axis, which contains all the even orders in its Fourier expansion:
\begin{equation}
    n = n_0 + n_2\cos2\phi + n_4\cos4\phi + \cdots\label{eq:pol fit}
\end{equation}
where $\phi$ is the in-plane polarization angle. For NIR pump conditions, we found that including up to the fourth order is enough to fit the data.
\clearpage

\section{Measurement of pulse duration}
\label{sec:pulse duration}
To confirm the constant pulse duration in each photon energy dependence measurement, we characterized both coherent responses during the pump-probe overlap as an \emph{in-situ}  approach and sum frequency generation (SFG) in a nonlinear crystal as a complementary \emph{ex-situ} method. This step is critical to conclusions in the main text because multiphoton and tunneling absorptions are highly nonlinear optical processes and therefore sensitively depend on the peak electric field $E$. Moreover, because laser pulses are Gaussian in experiments, different pulse durations with the same $E$ also give rise to distinct photocarrier densities. Thus, in order to make comparisons of datasets among multiple photon energies valid, it is imperative to ensure their temporal constancy.

The coherent optical response during light-matter interaction, sometimes referred to as a `coherent artifact', serves as an ideal \emph{in-situ} tool to quantify the pulse duration in transient reflectivity experiments. It happens when both the pump and the probe pulses are overlapped on the material target, which generates a nonlinear polarization that oscillates at the probe frequency:
\begin{equation}
    P(\omega_{\rm{probe}}) = \chi^{(3)}(\omega_{\rm{probe}};\omega_{\rm{pump}},-\omega_{\rm{pump}},\omega_{\rm{probe}})|E_{\rm{pump}}|^2E_{\rm{probe}}.
\end{equation}
The resulting signal then has a full width at half maximum (FWHM) $\tau=\sqrt{\tau_{\rm{pump}}^2+\tau_{\rm{probe}}^2}$ and exhibits a quadratic dependence on $E_{\rm{pump}}$.

\begin{figure}[!htb]
\includegraphics{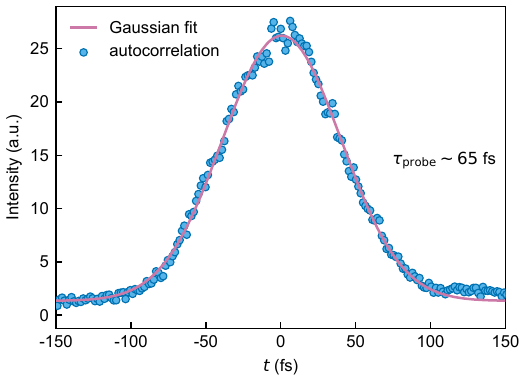}
\caption{\label{fig:autocorrelation} Probe pulse characterization. Autocorrelation trace of the 800~nm pulse in a frequency-resolved optical gating setup.}
\end{figure}

We first measured the 800~nm probe pulse duration using the frequency-resolved optical gating method. Its autocorrelation trace is shown in Fig.~\ref{fig:autocorrelation}, which leads to the duration of 65~fs at the sample position.

\begin{figure}[!htb]
\includegraphics{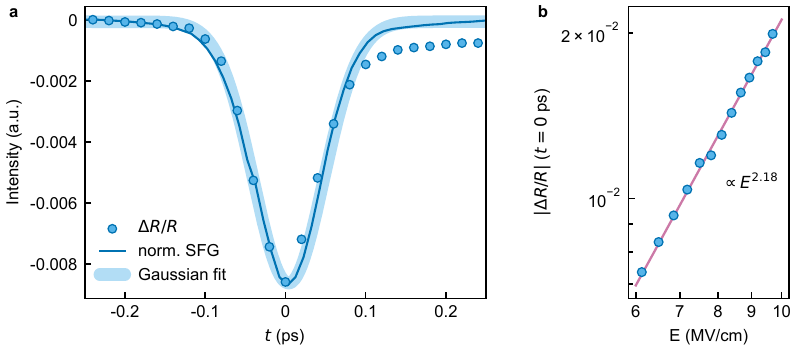}
\caption{\label{fig:coherent artifact overview} Pump pulse characterization through transient reflectivity. (a) Coherent signals in $\Delta R/R$ close to time zero (blue circles) with 0.9~eV pump. The normalized SFG signals with 800 nm probe are overlaid as a blue line. Transient reflectivity data on the rising edge is fitted using a Gaussian function and shown in a blue shade. (b) Electric field dependence of reflectivity transients at zero time delay (blue circles) and its linear regression (purple line) on a log-log scale.}
\end{figure}

Then, we characterized the coherent signal around zero time delay. Fig.~\ref{fig:coherent artifact overview}a shows the reflectivity transients with 0.9~eV pump. Both pulses are linearly polarized along the $a$ axis and the pump field is $E=10$~MV/cm. One can see that the signal is almost coherent on the rising edge and has a small residual background on the falling edge, corresponding to the incoherent electronic excitations discussed throughout the main text. We fitted the $\Delta R/R$ data on the rising edge with a Gaussian function and overlaid it in the blue shade. The resulting FWHM $\tau=100$~fs then leads to $\tau_{\rm{pump}}\approx75$~fs. In order to confirm that the signal originates from the cross-phase modulation effect, we also measured the dependence of the signal amplitude at $t=0$~ps on the pump electric field. A power-law fitting $|\Delta R/R|\propto E^{2\alpha}$ to the data gives rise to $\alpha\sim1$ (Fig.~\ref{fig:coherent artifact overview}b), confirming our hypothesis. Moreover, we also show the normalized cross-correlation signal between two pulses in a $\beta-$barium borate crystal at the sample position (Fig.~\ref{fig:coherent artifact overview}a, blue line), which overlaps with $\Delta R/R$ and the Gaussian fit exceptionally well, reinforcing its validity as an \emph{ex-situ} complementary tool to characterize $\tau_{\rm{pump}}$.

\begin{figure}[!htb]
\includegraphics[width=\linewidth]{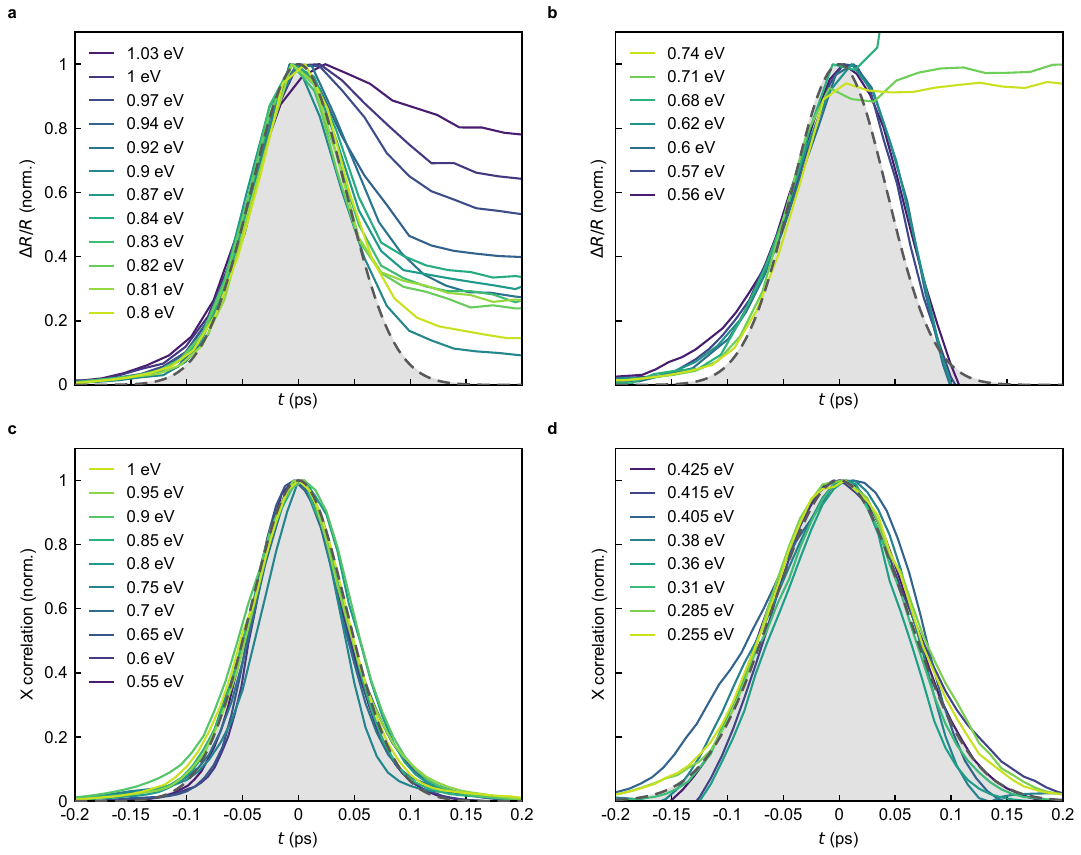}
\caption{\label{fig:pulse duration} Pump pulse characterization through sum frequency generation. (a) Normalized $\Delta R/R$ across the two-photon (a) and three-photon resonances (b). A Gaussian function with FWHM of 100~fs is overlaid on top of the data to highlight their overlap on the rising edge (black dashed line). (c) Normalized cross-correlation signals of 800 nm probe with NIR (c) and MIR pumps (d). A Gaussian function with FWHM of 100~fs/140~fs in (c)/(d) is overlaid to highlight their consistency with the data (black dashed line).}
\end{figure}

Having established the consistency of both approaches, we next utilized them to verify the constant pump pulse duration in the photon energy dependence measurements. We first scanned $\Delta R/R$ across the two-photon (Fig.~\ref{fig:pulse duration}a) and three-photon (Fig.~\ref{fig:pulse duration}b) transitions, and found that the rising edges of reflectivity transients all overlap well with a Gaussian function assuming $\tau_{\rm{pump}}=75$~fs. This is supported by the SFG signals across the same wavelength range (Fig.~\ref{fig:pulse duration}c), which also gives rise to $\tau_{\rm{pump}}=75$~fs for NIR pulses. Finally, we deployed the same approach to characterize the MIR pulse widths, where SFG can now be fitted by a Gaussian function with a FWHM of 140~fs (Fig.~\ref{fig:pulse duration}d). Deconvolving the probe pulse width, this leads to MIR pump duration $\tau_{\rm{pump}}=125$~fs.

\clearpage
\section{Experimental probes of the photoexcitation yield}
\label{sec:probe}
Under the dilute excitation conditions of this work, the transient differential reflectivity $\Delta R/R$ and the coherent acoustic phonon (CAP) amplitude provide complementary and reliable relative probes of the photoexcitation yield. Their conversion to an absolute carrier density is not assumed to be universal, because the response generated per photoexcited pair can depend on the carrier momentum distribution and excitation-induced changes in the electronic structure and the complex dielectric function. We show below that these possible weightings do not materially alter the relative field and photon-energy dependences analyzed here. This conclusion is supported independently by semiconductor Bloch equations (SBE) simulations and by the microscopic deformation-potential calculation for a driven Hubbard model in~\cite{andrei2025subgap}.

\subsection{Transient differential reflectivity at 500~fs}
The transient differential reflectivity $\Delta R/R$ used in the main text is evaluated at a pump--probe delay of 500~fs, well beyond temporal overlap with the pump. It therefore probes a persistent photoinduced dielectric response rather than the instantaneous optical coherence or field dressing. By this delay, the optical coherence has decayed and the carriers are expected to have undergone substantial intraband redistribution, before interband recombination takes place.

Treating $n$ as a number, the differential reflectivity can be expanded in $n$ as
\begin{equation}
\frac{\Delta R}{R_0}(n)
\equiv \frac{R(n)-R_0}{R_0}
=\frac{n}{R_0}\left.\frac{\partial R}{\partial n}\right|_{n=0}
+\mathcal{O}(n^2),
\label{eq:r25_reflectivity_linear}
\end{equation}
where $R_0=R(0)$. The small measured response (Figs.~\ref{fig:frequency dependence}a,b) and the expected dilute excitation are consistent with retaining the leading linear term (provided that the coefficient $\frac{\partial R}{\partial n}|_{n=0}$ is generally nonzero), so \(\Delta R/R\) serves as a relative population-sensitive proxy.

\subsection{SBE simulations}
To test explicitly how momentum redistribution affects an energy-weighted observable, we performed complementary SBE simulations \cite{haug2009quantum,kira2012semiconductor}. The SBE evolve the interband polarization $P_q$ and conduction-band population $n_q$ in the Houston frame,
\begin{align}
\dot P_q &=-i\left[
\frac{\Omega(k(t))}{\hbar}-\frac{i}{T_2}\right]P_q
-\frac{i}{\hbar}d_{cv}E(t)\left(1-2n_q\right),
\label{eq:si_r25_sbe_polarization}\\
\dot n_q &=-\frac{2}{\hbar}\operatorname{Im}
\!\left\{[d_{cv}E(t)]^{*}P_q\right\},
\label{eq:si_r25_sbe_population}
\end{align}
where $k(t)=q+eA(t)/\hbar$, $E(t)=-\partial_tA(t)$, $d_{cv}$ is the interband dipole matrix element, and $\Omega(k)=E_c(k)-E_v(k)$ is the momentum-dependent gap. This formulation treats nonlinear interband absorption and intraband field acceleration within the same calculation. We used the collisionless limit $T_2^{-1}=0$ and included no population-relaxation term, thereby isolating the coherent injection process and its residual post-pulse distribution.

\begin{figure}[htbp]
    \centering
    \includegraphics{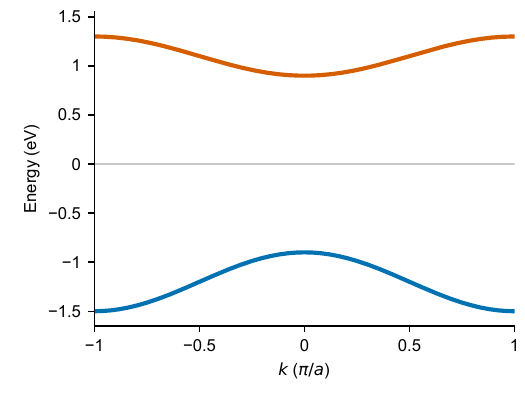}
    \caption{Two-band dispersion used in the SBE simulations. The upper orange and lower blue curves denote the conduction and valence bands, respectively. The model has a direct gap $\Delta=1.8$~eV at $k=0$, a conduction-band width of $0.4$~eV, and a valence-band width of $0.6$~eV. The gray horizontal line denotes the chosen zero of energy.}
    \label{fig:si_r25_sbe_bands}
\end{figure}

The one-dimensional two-band model was chosen as
\begin{align}
E_c(k)&=\frac{\Delta}{2}+\frac{W_c}{2}\left(1-\cos ka\right),\\
E_v(k)&=-\frac{\Delta}{2}-\frac{W_v}{2}\left(1-\cos ka\right),
\label{eq:si_r25_sbe_bands}
\end{align}
with $\Delta=1.8$~eV, $W_c=0.4$~eV, and $W_v=0.6$~eV (Fig.~\ref{fig:si_r25_sbe_bands}). These bandwidths are representative of the relatively narrow near-edge manifolds in calculated bulk NiPS$_3$ dispersions~\cite{lane2020thickness}. This model is intended to test the relation between carrier number and excitation energy for the relevant gap-to-bandwidth ratio, rather than to reproduce the full multiband and anisotropic electronic structure of NiPS$_3$. We used a constant dipole matrix element and Gaussian pulses with a 100~fs intensity full width at half maximum for excitation.

From the residual post-pulse distribution, we calculate the total density of excited pairs and their total electronic excitation energy,
\begin{equation}
N=\sum_k n_k,
\qquad
\delta E_{\mathrm{exc}}=\sum_k\Omega(k)n_k.
\label{eq:si_r25_sbe_observables}
\end{equation}
In the perturbative multiphoton regime, carriers are concentrated near the resonant momenta satisfying $\Omega(k)=m\hbar\omega$, where $m$ is the photon order. It follows that
\begin{equation}
\delta E_{\mathrm{exc}}\simeq m\hbar\omega N,
\label{eq:si_r25_multiphoton_energy}
\end{equation}
so the carrier density and total excitation energy have essentially the same field dependence even though the distribution is localized at specific momenta.

\begin{figure}[htbp]
    \centering
    \includegraphics[width=\textwidth]{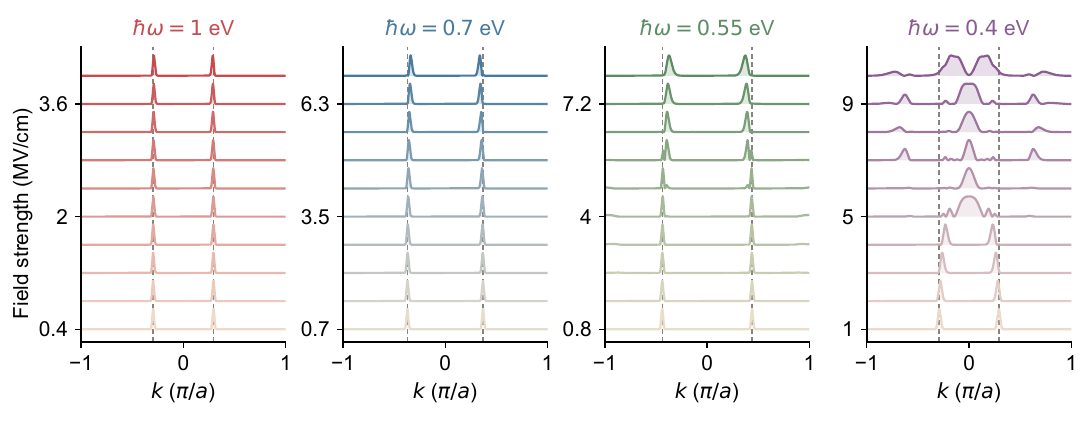}
    \caption{Momentum-resolved carrier distributions from the SBE calculation. Residual conduction-band occupations $n_k$ after photoexcitation are shown for four photon energies as a function of the peak field. Each trace is normalized to its own maximum and vertically offset for clarity. Vertical dashed lines mark the bare-band resonance momenta satisfying $E_c(k)-E_v(k)=m\hbar\omega$, with $m=2$, 3, 4, and 5 for $\hbar\omega=1$, $0.7$, $0.55$, and $0.4$~eV, respectively.}
    \label{fig:si_r25_sbe_distributions}
\end{figure}

Figure~\ref{fig:si_r25_sbe_distributions} shows the calculated momentum distributions. At $\hbar\omega=1$ and $0.7$~eV, the excitation remains concentrated near the two- and three-photon resonant momenta over the displayed field ranges. At $0.55$~eV, the distribution initially follows the four-photon resonance and develops additional structure with increasing field. This redistribution becomes more pronounced at $0.4$~eV, where several features coexist in the momentum space. Thus, the nonadiabatic strong-field response is still momentum-dependent, and the distribution retains band- and channel-specific structure even while its field dependence becomes nonperturbative.

\begin{figure}[htbp]
    \centering
    \includegraphics[width=\textwidth]{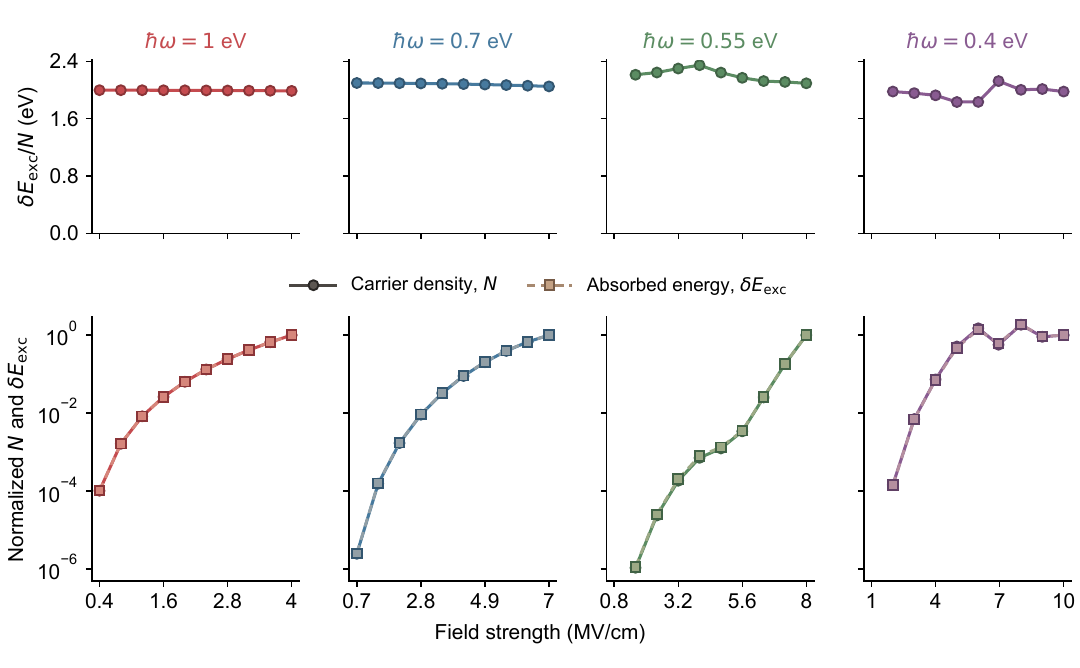}
    \caption{Comparison of total carrier density and electronic excitation energy. The upper row shows the mean excitation energy per carrier, $\delta E_{\mathrm{exc}}/N$, versus the peak field. The lower row compares the total residual carrier density $N$ and excitation energy $\delta E_{\mathrm{exc}}$ on a logarithmic scale. Within each panel, the two quantities are normalized independently to their values at the highest simulated field.}
    \label{fig:si_r25_sbe_totals}
\end{figure}

Because the relevant valence and conduction manifolds are narrow compared with the $1.8$~eV charge-transfer gap, redistribution within those manifolds changes the mean energy per pair by much less than the many-orders-of-magnitude change in $N$, which is directly confirmed in Fig.~\ref{fig:si_r25_sbe_totals}. The independently normalized field dependences of $N$ and $\delta E_{\mathrm{exc}}$ nearly coincide for all four photon energies. Meanwhile, $\delta E_{\mathrm{exc}}/N$ remains between approximately $1.8$ and $2.35$~eV and varies by only a few tenths of an electronvolt within each scan. Under these excitation conditions, an observable weighted by deposited electronic energy therefore retains essentially the same relative field dependence as the carrier yield, although its conversion factor may not be universal.

\subsection{Microscopic relation between photoexcitation and CAP generation}
The measured ${\sim}27$-GHz oscillation is a coherent longitudinal-acoustic strain wave. Because the pump field decreases with depth, photoexcitation produces a spatially nonuniform change in electronic bond order and hence a deformation-potential stress. Ref.~\cite{andrei2025subgap} describes this process microscopically using a Su-Schrieffer-Heeger-type coupling between the electronic kinetic energy and the lattice displacement. Assuming the separation of timescales following photoexcitation
\begin{equation}
T_{\mathrm{pump}}\ll\tau_{\mathrm{intraband}}
\ll\tau_{\mathrm{CAP}}\ll\tau_{\mathrm{recomb}},
\label{eq:si_r25_cap_timescales}
\end{equation}
the number of doublon--holon pairs is approximately conserved while intraband scattering redistributes them within the upper and lower Hubbard bands. The resulting post-relaxation electronic state acts as a rapid quench on the much slower acoustic mode.

For a mode of wavevector $q$, the displacement is proportional to the corresponding depth Fourier component of the pump-induced kinetic-energy change,
\begin{equation}
A_{\mathrm{CAP}}(q)\propto
\left|\int dz\,
\delta\!\left\langle E_{\mathrm{kin}}^{\mathrm{final}}(z)\right\rangle
e^{-iqz}\right|.
\label{eq:si_r25_cap_kinetic_energy}
\end{equation}
The ${\sim}37$~ps acoustic period is much longer than the pump and intraband-redistribution times, making this source effectively displacive. The nearly cosine phase measured experimentally, $\varphi=-0.14\pi$ (Sec.~\ref{sec:fit trr}), is consistent with such a displacive launch mechanism.

\begin{figure}[htbp]
    \centering
    \includegraphics[width=\textwidth]{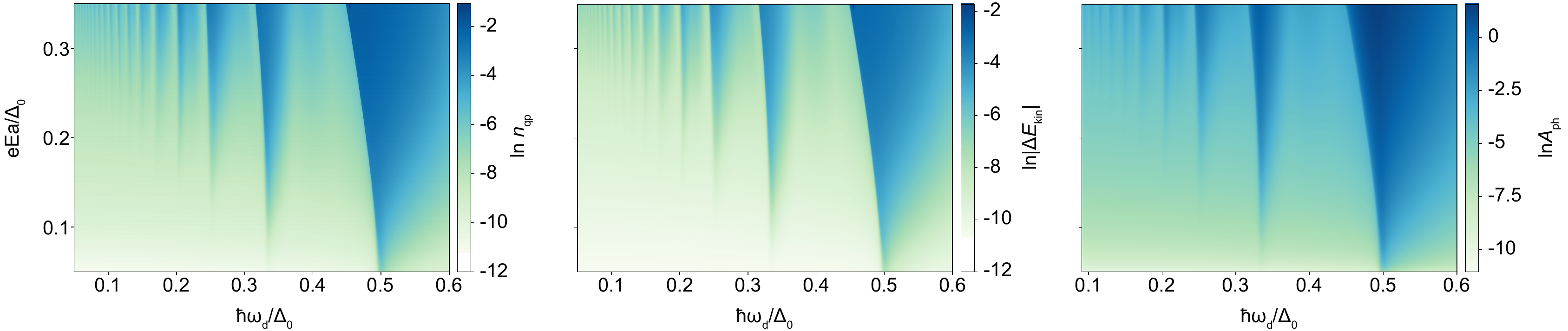}
    \caption{Microscopic relation between photoexcitation and CAP generation. Left, photoexcited doublon-holon density $n_{\mathrm{qp}}$ immediately after the pump. Center, magnitude of the pump-induced change in the post-relaxation local kinetic-energy density, $|\Delta E_{\mathrm{kin}}|$. Right, amplitude of the induced acoustic displacement $A_{\mathrm{ph}}$ at the wavevector $q=\lambda_{\mathrm{pump}}^{-1}$ selected by the pump penetration depth. All color scales are logarithmic. The panels are adapted from Fig.~2a and Figs.~10a,b in~\cite{andrei2025subgap}.}
    \label{fig:si_r25_hubbard_cap}
\end{figure}

Figure~\ref{fig:si_r25_hubbard_cap} compares the independently calculated photoexcited doublon-holon density (left), the post-relaxation kinetic-energy change that generates the stress (middle), and the resulting acoustic displacement (right). The kinetic-energy change and acoustic displacement are calculated from the relaxed driven state. The same pump-frequency thresholds and field-dependent structures are transferred from $n_{\mathrm{qp}}$ to $|\Delta E_{\mathrm{kin}}|$ and then to $A_{\mathrm{ph}}$. At sufficiently high density, changes in the intraband distribution can alter the kinetic energy and stress per pair, so exact proportionality is not expected. Over most of the parameter range relevant here, however, this weighting introduces only secondary differences and preserves the threshold landscape and relative field dependence.

\subsection{Other possible contributions}
The carrier-mediated displacive mechanism above is distinct from ordinary impulsive stimulated Raman scattering (ISRS) \cite{yan1985impulsive}, whose lowest-order force is
\begin{equation}
F_{\mathrm{ISRS}}(t)\propto
\left.\frac{\partial\chi}{\partial Q}\right|_{Q=0}|E(t)|^2.
\label{eq:si_r25_isrs_force}
\end{equation}
Ordinary ISRS does not require a residual carrier population and therefore gives an amplitude proportional to $E^2$. By contrast, the CAP amplitude follows $E^4$ at $0.97$~eV and $E^6$ at $0.71$~eV, consistent with two- and three-photon carrier generation. Such a scaling evolves into the tunneling behavior across $\gamma_K\simeq1$. Therefore, we find no positive evidence that such a contribution dominates the measured response.

The same considerations constrain possible defect-assisted backgrounds as well. Defect-assisted absorption or trapping may change the stress per excitation, acoustic damping, or optical detection coefficient and thereby alter the response prefactor. However, a dominant defect-only mechanism still fails to explain simultaneously the multiphoton and tunneling field dependences, the multiple intrinsic-gap-related thresholds, and their systematic strong-field displacement.

\bigskip
Taken together, the dilute-response expansion, the SBE comparison between $N$ and $\delta E_{\mathrm{exc}}$, and the independent deformation-potential calculation support using $\Delta R/R$ and the CAP amplitude as complementary relative measures of the photoexcitation yield under our current experimental conditions. Energy-, momentum-, and transduction-dependent weighting precludes an absolute carrier-density calibration and can modify the response prefactor. It does not, however, account for the observed multiphoton(-like) thresholds, their field-induced evolution, or the crossover from perturbative power-law scaling to tunneling-like field dependence on which our conclusions are based.

\clearpage
\section{Fitting of transient reflectivity traces}
\label{sec:fit trr}
\begin{figure}[!htb]
\includegraphics{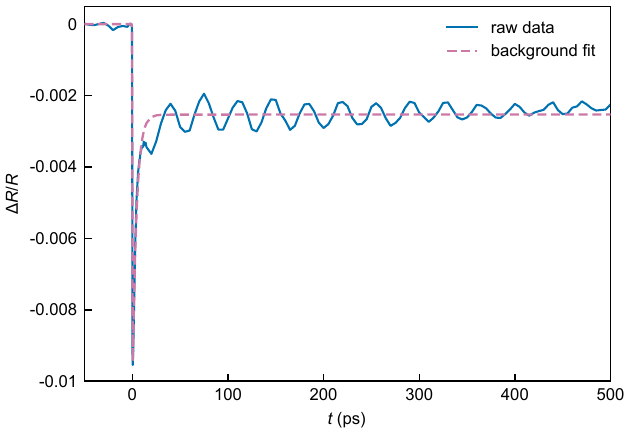}
\caption{\label{fig:fit} Fitting of differential reflectivity transients. 2 eV pump $\Delta R/R$ trace at $E = 1.9$ MV/cm. The purple dashed line represents the fitted background.}
\end{figure}

We summarize the fitting procedure to transient reflectivity traces presented in the main text here. Figure~\ref{fig:fit} shows the same data from Fig.~\ref{fig:m1}b. We fit the background by a bi-exponential decay function plus a constant offset:
\begin{equation}
    \frac{\Delta R}{R} = a_1e^{-\frac{t}{\tau_1}} + a_2e^{-\frac{t}{\tau_2}} + c.
\end{equation}
The fitted background is shown as the purple dashed line in Fig.~\ref{fig:fit}, which describes the incoherent part reasonably well. The residual acoustic phonon contribution in the left inset of Fig.~\ref{fig:m1}b is then fitted by a damped cosine function:
\begin{equation}
    \frac{\Delta R}{R}|_{\rm{phonon}} = a_3e^{-\frac{t}{\tau_3}}\cos(\Omega t+\varphi).
\end{equation}
The determined phase $\varphi=-0.14\pi$ is close to 0, supporting its displacive excitation mechanism.

The coherent phonon data presented in the other parts of the main text is processed in the same way.

\clearpage
\section{Pump energy-, field-, and polarization-dependent datasets}
\label{sec:datasets}
\begin{figure}[!htb]
\includegraphics[width=\textwidth]{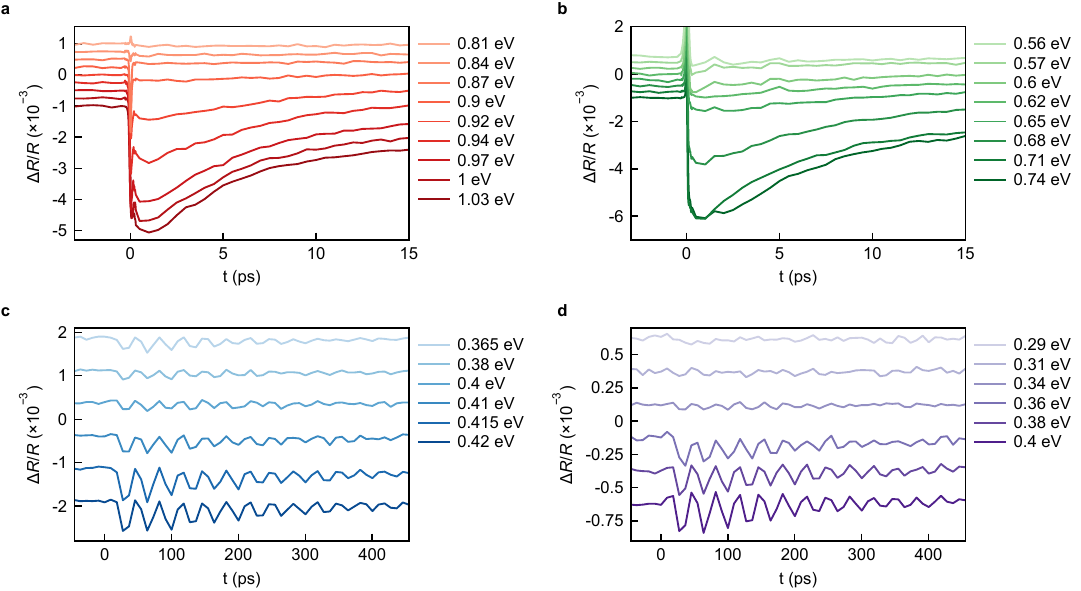}
\caption{$\Delta R/R$ at different photon energies. Pump photon energy dependence of the reflectivity transients across the two-photon (a), three-photon (b), five-photon (c), and six-photon resonances (d). The absence of pronounced responses close to time-zero in the MIR pump reflectivity transients (c and d) is due to different sampling in the time domain. Curves are vertically offset for clarity.}
\label{fig:frequency dependence}
\end{figure}

We present the complete reflectivity transients used in Figs.~\ref{fig:m2} and \ref{fig:m3} here. Figure~\ref{fig:frequency dependence} shows the pump photon energy dependence of $\Delta R/R$ across the two-photon (Figs.~\ref{fig:frequency dependence}a and~\ref{fig:m2}e), three-photon (Figs.~\ref{fig:frequency dependence}b and Fig.~\ref{fig:m3}a), five-photon (Figs.~\ref{fig:frequency dependence}c and Fig.~\ref{fig:m3}d), and six-photon (Figs.~\ref{fig:frequency dependence}d and Fig.~\ref{fig:m3}g) resonances, respectively. In each case, we can clearly see that the response rapidly grows as $\hbar\omega$ goes across the corresponding threshold $\tilde\Delta/n$. 

\begin{figure}[!htb]
\includegraphics[width=\textwidth]{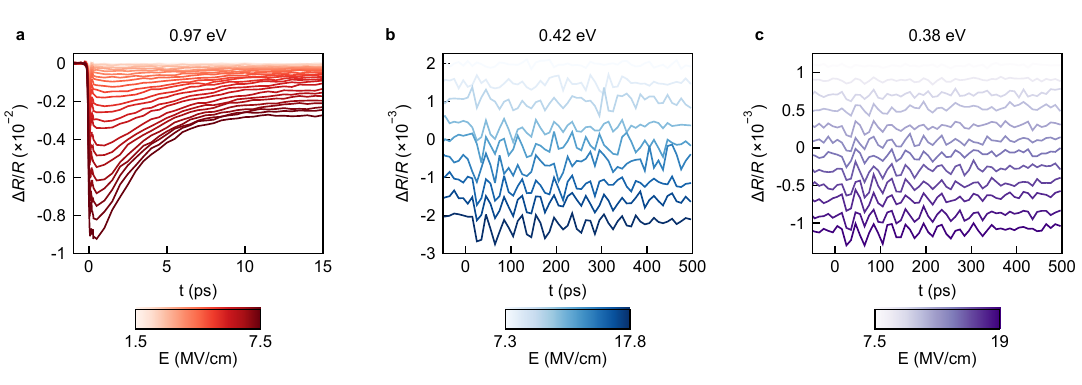}
\caption{$\Delta R/R$ at different field strengths. Pump electric field dependence of the reflectivity transients at $\hbar\omega$ = 0.97 eV (a), 0.42 eV (b), and 0.38 eV (c). The absence of pronounced responses close to time-zero in the MIR pump reflectivity transients (b and c) is due to different sampling in the time domain. Curves in (b) and (c) are vertically offset for clarity.}
\label{fig:f dep}
\end{figure}

Figure~\ref{fig:f dep} shows the pump electric field dependence of $\Delta R/R$ at 0.97~eV (a; Fig.~\ref{fig:m2}c), 0.42~eV (b; Fig.~\ref{fig:m3}f), and 0.38~eV (c; Fig.~\ref{fig:m3}i) pump, respectively. For the field-dependent cuts in main-text Figs.~\ref{fig:m3}c,f,i, the measured photoexcitation proxy $S$ was compared with the tunneling-like and fixed-order multiphoton forms
\begin{align}
S_{\mathrm{tun}}(E_0)&=A E_0^2
\exp\!\left(-\frac{E_{\mathrm{th,eff}}}{E_0}\right),
\label{eq:si_r22_tun_fit}\\
S_{N\mathrm{ph}}(E_0)&=B_N E_0^{2N},
\label{eq:si_r22_mp_fit}
\end{align}
For the dashed curves, $N$ was fixed to 4, 5, and 6 at $\hbar\omega=0.56$, 0.42, and 0.38~eV, respectively, and only $B_N$ was varied. For the solid curves, $A$ and $E_{\mathrm{th,eff}}$ were varied independently at each photon energy; no universal threshold field was imposed. The resulting effective field scales are $E_{\mathrm{th,eff}}=6.57$, 7.53, and 9.02~MV/cm, respectively. These values are close to the field scales reported near the Keldysh crossover in a lead-halide perovskite with a slightly larger gap of approximately 2.3~eV~\cite{zhang2025keldysh}.

The fitted values should be interpreted as effective inverse-field curvature scales rather than precise microscopic onset fields. The detected response is integrated over the Gaussian spatial and temporal profile of the pump, and the average of $E^2\exp(-E_{\mathrm{th}}/E)$ cannot be represented by the same homogeneous-field expression with an unchanged $E_{\mathrm{th}}$. The explicit calculations in Sec.~\ref{sec:Keldysh parameter} show that fitting the integrated response can shift the recovered exponential field scale by factors of order unity. Therefore, the more principal implication of $E_{\mathrm{th,eff}}$ is the robust departure of the measured field dependences from the nominal perturbative laws $E_0^8$, $E_0^{10}$, and $E_0^{12}$.

In addition, we note that such a frequency-dependent effective tunneling threshold aligns well with our theory (Secs.~\ref{sec:Hubbard} and \ref{sec:Keldysh theory}) and existing experimental literature. In solids, the crossover from frequency-independent quasi-static tunneling to frequency-dependent tunneling has been observed for deep centers in semiconductors~\cite{ganichev1998carrier,ganichev2002tunnelling}; related frequency-dependent effective ionization and saturation scales have also been reported near the strong-field crossover in atomic and molecular systems~\cite{lai2017experimental,lezius2001nonadiabatic,lezius2002polyatomic}.

\begin{figure}[!htb]
\includegraphics{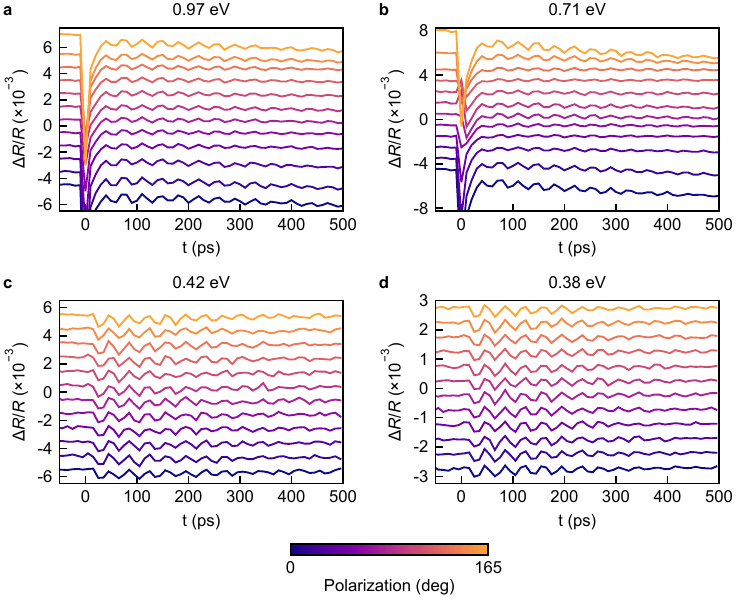}
\caption{$\Delta R/R$ at different in-plane polarization angles. Phonon dynamics at different in-plane electric field pump polarization angles for $\hbar\omega$ = 0.97 eV (a), 0.71 eV (b), 0.42 eV (c), and 0.38 eV (d). The absence of pronounced responses close to time-zero in the MIR pump reflectivity transients (c and d) is due to different sampling in the time domain. Curves are vertically offset for clarity.}
\label{fig:pol dep}
\end{figure}

Figure~\ref{fig:pol dep} shows the pump polarization dependence of $\Delta R/R$ at 0.97 eV (a; Fig.~\ref{fig:m2}e inset), 0.71 eV (b; Fig.~\ref{fig:m2}e inset), 0.42 eV (c; Fig.~\ref{fig:m3}e), and 0.38 eV (d; Fig.~\ref{fig:m3}h) pump, respectively. In the multiphoton regime, $n$ exhibits a pronounced anisotropy with the maximal carrier generation efficiency at 0$\degree$; upon entering the tunneling regime, this anisotropy diminishes and $n$ becomes increasingly isotropic. The reduction of the angular contrast can be understood from the orientation-dependent Zener tunneling probability. The excitation rate can be written as \cite{zener1934theory}
\begin{equation}
W(\phi,E)\propto
\exp\!\left[-\frac{(2\Delta)^{3/2}\sqrt{m^*(\phi)}}
{3e\hbar E}\right]
\equiv \exp\!\left[-\frac{B(\phi)}{E}\right],
\label{eq:si_r23_orientation_rate}
\end{equation}
where $m^*(\phi)$ is the reduced electron-hole mass along the field direction. The rate ratio for two in-plane orientations therefore obeys
\begin{equation}
\frac{W(\phi_1,E)}{W(\phi_2,E)}\sim\exp\!\left[-\frac{B(\phi_1)-B(\phi_2)}{E}\right].
\label{eq:si_r23_orientation_ratio}
\end{equation}
The full excitation rate can also contain prefactors associated with transition matrix elements, dephasing, and field strength, which generally vary much more slowly than the exponential and therefore provide secondary corrections to the leading orientation dependence. Thus, although unequal effective masses produce an exponential orientation dependence, their exponential contribution to the logarithmic contrast decreases as $1/E$ and the response will become less anisotropic as the field increases. A similar suppression of the anisotropic nonlinear absorption with increasing field was also observed in diamond, where the carrier population ratio for excitation along $\langle110\rangle$ and $\langle100\rangle$ decreased toward unity on approaching the tunneling regime~\cite{kozak2019anisotropy}.

We quantified the experimental sensitivity to such a residual modulation directly from the angular datasets in Figs.~\ref{fig:m3}b,e,h. The independent angular points were fitted to the $2/m$-symmetric form
\begin{equation}
S(\phi)=S_0+S_2\cos2\phi
+S_4\cos4\phi,
\label{eq:si_r23_angular_fit}
\end{equation}
and the empirical one-standard-deviation noise from the residuals is defined as
\begin{equation}
\sigma_{\mathrm{noise}}
=\left[\frac{1}{N-3}\sum_{i=1}^{N}
\bigl(S_i-S_{\mathrm{fit},i}\bigr)^2\right]^{1/2}.
\label{eq:si_r23_noise}
\end{equation}

The results are summarized in Table~\ref{tab:polarization_noise}. The \(0.56\)~eV response exhibits clearly resolved twofold and fourfold anisotropy. By contrast, neither harmonic at \(0.42\) or \(0.38\) eV exceeds twice its fitted uncertainty. Using an approximate \(2\sigma\) detection criterion, our measurements therefore cannot reliably resolve harmonic modulations smaller than about \(11.2\%\) of the mean signal at \(0.42\) eV or \(14.5\%\) at \(0.38\) eV. Equivalently, the point-to-point scatter amounts to \(13.7\%\) and \(17.7\%\), respectively. We therefore describe the low-photon-energy response as isotropic within experimental resolution. This conclusion does not exclude a weaker residual anisotropy below our sensitivity, but demonstrates that the pronounced angular modulation observed at 0.97 and 0.71 eV (Fig.~\ref{fig:m2}e inset) is strongly suppressed upon entering the nonperturbative tunneling regime.

\begin{table}[htbp]
\centering
\caption{\textbf{Polarization-fit parameters and experimental noise levels.}
Here, the uncertainties of $S_2$ and $S_4$ represent one standard deviation.}
\label{tab:polarization_noise}
\small
\begin{tabular}{@{}lccccc@{}}
\toprule
Panel & $\hbar\omega$ & $S_0$ &
$\sigma_{\mathrm{noise}}/S_0$ &
$|S_2|/S_0$ & $|S_4|/S_0$ \\
 & (eV) & (a.u.) & (\%) & (\%) & (\%) \\
\midrule
Fig.~\ref{fig:m3}b & $0.56$ & $1.262$ & $4.1$  & $17.0\pm1.4$ & $16.1\pm1.4$ \\
Fig.~\ref{fig:m3}e & $0.42$ & $6.280$ & $13.7$ & $3.2\pm5.6$  & $1.3\pm5.6$ \\
Fig.~\ref{fig:m3}h & $0.38$ & $1.729$ & $17.7$ & $1.0\pm7.2$  & $6.8\pm7.2$ \\
\bottomrule
\end{tabular}
\end{table}

\clearpage
\section{Temperature dependence of nonlinear absorption in NiPS$_3$}
\label{sec:t dep}
The equilibrium optical conductivity of NiPS$_3$ rises from the charge-transfer onset near $\Delta=1.8$~eV and reaches a pronounced maximum near 2.22~eV \cite{kim2018charge}. Temperature-dependent optical spectra show that the magnetically enhanced spectral weight associated with the 2.22~eV structure is substantially reduced above the N\'eel temperature $T_N$~\cite{kim2018charge}. This provides a useful test of whether this feature is required for the nonlinear response observed under sub-gap excitation.

\begin{figure}[htb]
\centering
\includegraphics{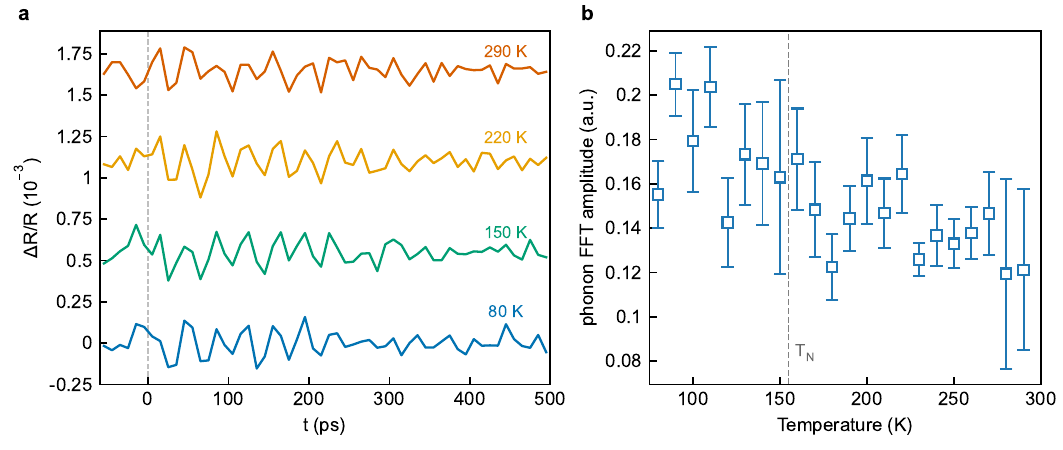}
\caption{Temperature dependence of the nonlinear CAP response. (a) Representative $\Delta R/R$ traces measured at 80, 150, 220, and 290 K at fixed $\hbar\omega=0.38$~eV excitation with 18~MV/cm incident peak field. An incoherent background was subtracted, and the traces are vertically offset for clarity. (b) CAP FFT amplitude as a function of temperature. Error bars represent the one-standard-deviation uncertainty propagated from the off-peak FFT noise. The vertical dashed line marks $T_N$.}
\label{sfig:t dep}
\end{figure}

To this end, we measured the CAP response at $\hbar\omega=0.38$~eV and 18~MV/cm while varying the sample temperature from 80 to 290~K. As shown in Fig.~\ref{sfig:t dep}, the absorption signal remains clearly resolved above $T_N\simeq155$~K, with no discontinuous loss on crossing the magnetic transition within the spectral noise uncertainty. This implies that the charge dynamics is not controlled exclusively by the magnetically enhanced 2.22~eV structure.

We next consider how these two features of the charge-transfer spectrum enter the resonance structure. For the excitation conditions used in Fig.~\ref{fig:m3}, $N\hbar\omega=\Delta+E_p$ gives resonance energies of approximately 0.618, 0.402, and 0.352~eV for $N=3$, 5, and 6, respectively, in agreement with the observed steps near 0.62, 0.4, and 0.35~eV. By contrast, using the corresponding ponderomotive energies $E_p=0.054$, 0.21, and 0.312~eV, resonances terminating at the 2.22~eV maximum would occur at
\begin{equation}
\frac{2.22+0.054}{3}=0.758~\mathrm{eV},\qquad
\frac{2.22+0.21}{5}=0.486~\mathrm{eV},\qquad
\frac{2.22+0.312}{6}=0.422~\mathrm{eV}.
\label{eq:supp_peak_resonances}
\end{equation}
All three values lie above the corresponding experimental steps. The four-photon resonance of the 2.22~eV maximum under the conditions of Fig.~\ref{fig:m3}a is expected near $(2.22+0.054)/4=0.569$~eV. Although it may contribute near the lower boundary of that scan, it lies approximately 50~meV below the observed 0.62~eV step and cannot account for its position. These comparisons support using the dressed 1.8~eV onset as the primary scale for the step energies while allowing the 2.22~eV maximum to modify the response amplitude and fine structure.

The influence of equilibrium spectral structure differs between the perturbative multiphoton and tunneling limits. Multiphoton excitation is energy- and momentum-selective and is therefore sensitive to the joint density of states and optical matrix elements that produce fine structure in the equilibrium optical conductivity. Landau-Zener tunneling is instead local in real space and, in the adiabatic limit, produces a broad and comparatively weakly momentum-selective distribution of doublon-holon pairs~\cite{zener1934theory,oka2012nonlinear}. In the finite-frequency nonadiabatic regime relevant here, however, tunneling-like field dependence coexists with multiphoton-like photon-energy resonances, allowing spectral fine structure to remain visible without setting the principal threshold.

Different nominal photon-order channels can also overlap in this regime. For a mid-infrared drive in Fig.~\ref{sfig:t dep}, $E_p\approx0.237$~eV. The nominal six-photon resonance of the 2.22~eV maximum and the five-photon resonance of the dressed onset are then
\begin{equation}
\frac{2.22+E_p}{6}=0.4095~\mathrm{eV},\qquad
\frac{1.8+E_p}{5}=0.4074~\mathrm{eV}.
\label{eq:supp_overlapping_channels}
\end{equation}
Their separation is approximately 2~meV, much smaller than the typical $\sim15$~meV linewidth of the MIR pulses. A structure near 0.41~eV can therefore contain contributions associated with both parts of the same charge-transfer manifold. Taken together with the temperature dependence, this overlap explains how the 2.22~eV maximum may influence the fine structure and magnitude of a response whose principal step positions are governed by the dressed onset of interband transition.

\clearpage
\section{Electric field dependence of reflectivity transients at different polarization angles}
\label{sec:pol dep e dep}
\begin{figure}[!htb]
\includegraphics[width=\textwidth]{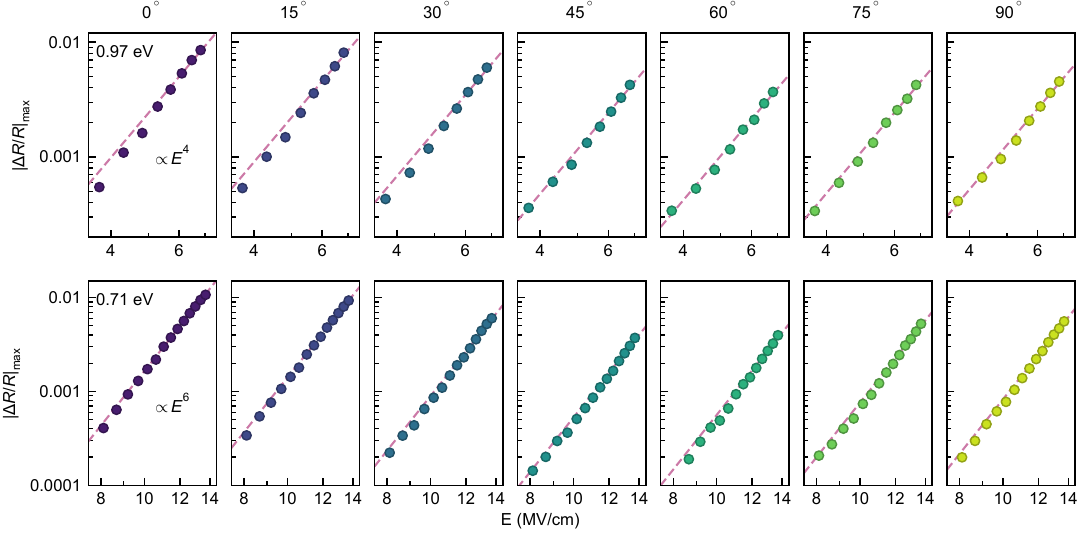}
\caption{\label{fig:field dependence 2 ph 3 ph} Field scaling of $\Delta R/R$ at different in-plane polarization angles. Electric field dependence of differential transient reflectivity at different polarization angles with 0.97~eV (top row) and~0.71 eV pump (bottom row). The pink dashed line in each panel shows the power-law fitting to the corresponding multiphoton process. $0^\circ$ and $90^\circ$ denote $a$ and $b$ axes, respectively.}
\end{figure}

We performed additional measurements to confirm the multiphoton nature of anisotropic charge excitations with 0.97~eV and 0.71~eV pump. Fig.~\ref{fig:field dependence 2 ph 3 ph} shows the electric field dependence of $|\Delta R/R|_{\rm{max}}$ at different polarization angles on a log-log scale. The precise alignment of power-law fitting with the experimental data points clearly indicates that the multiphoton processes are activated across all in-plane polarization angles.

\clearpage
\section{Possible role of oblique incidence geometry}
\label{sec:oblique incidence}
\begin{figure}[!htb]
\includegraphics{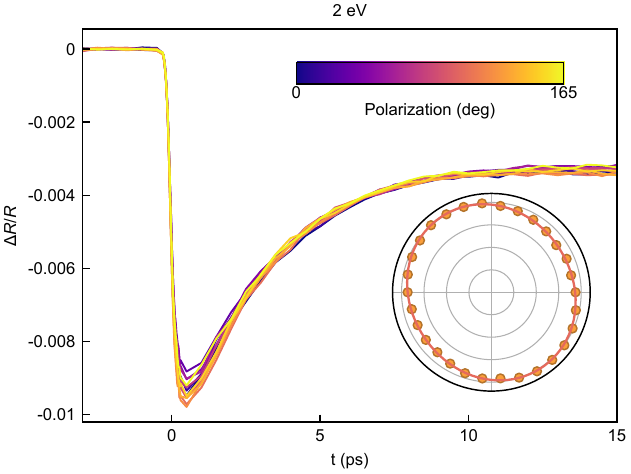}
\caption{\label{fig:oblique incidence} $\Delta R/R$ at different in-plane polarization angles. Differential reflectivity transients at different electric field polarization with 2~eV pump at 90~K. The inset shows the signal (orange circles) at $t=200$~fs overlaid by a fitting to $\cos{(2\phi+\phi_0)}$ (orange curve) where $\phi$ is the polarization angle and $\phi_0$ is the offset.}
\end{figure}

Although a small finite oblique incident angle $\sim 15^\circ$ is adopted in our experiments, we performed additional measurements to rule out its impact on the prominent polarization dependence observed with sub-gap pump conditions. Since the symmetry of charge excitations in this regime is not \emph{a priori} clear, we utilized an above-gap pump at 2~eV with $E=1.9$~MV/cm and examined the transient reflectivity at various linear polarization angles (Fig.~\ref{fig:oblique incidence}). $\Delta R/R$ traces appear almost identical in the short time delay window, and the maximum of its amplitude at $t=200$~fs as a function of polarization is shown in the inset of Fig.~\ref{fig:oblique incidence}. The data exhibits a slight asymmetry with a period of $\pi$, characteristic of responses from an oblique incidence configuration. We note that the linear dichroism at this photon energy below the N\'{e}el temperature \cite{hwangbo2021highly} also leads to slight asymmetry, and we estimate the contribution from our experimental geometry to be below 10\%. This is much smaller than the asymmetry observed in the multiphoton regime and lower than the noise level of signals in the tunneling regime. Additionally, an oblique incidence geometry only gives rise to a two-fold rotation symmetry, distinct from the multiphoton behaviors which include higher harmonic components ($\cos{4\phi}$).

\clearpage
\section{Definition of the Keldysh parameter}
\label{sec:Keldysh parameter}
Keldysh's original treatment to ionization in the field of a strong electromagnetic wave considered atomic ionization and interband excitation in crystals within the same strong-field framework \cite{keldysh1964ionization}. In the atomic problem, the field deforms the binding potential and promotes an electron across the ionization potential $I_p$ into the continuum, leaving a residual ion. In a crystal, the corresponding process in solids transfers an electron from an occupied to an unoccupied state and creates an electron-hole pair. Accordingly, $\gamma_K$ retains the same timescale interpretation: $\gamma_K\gg1$ is the multiphoton limit, $\gamma_K\ll1$ is the quasi-static tunneling limit, and $\gamma_K\sim1$ marks the nonadiabatic crossover. The atomic ponderomotive energy $U_p$ likewise corresponds, at the effective-mass level, to the combined electron--hole quiver energy $E_p$, which shifts the effective excitation threshold.

The Keldysh theory provides an analytical expression for the cycle-averaged rate of the interband transitions in a monochromatic electric field of arbitrary strength. The dimensionless Keldysh parameter determines the adiabaticity of the interband excitations
\begin{equation}\label{eq:gk}
    \gamma_K = \frac{\omega\sqrt{m\Delta}}{eE}
\end{equation}
where \emph{m} is the reduced mass of the electron and the hole, \emph{e} is the elementary charge, and $\Delta$ is the bandgap of the insulator \cite{keldysh1964ionization}. It should be noted that the derivation of $\gamma_K$ is based on a Kane's band model, which is oversimplified for realistic semiconductors and dielectrics. Essentially, $\gamma_K$ is the ratio of the light frequency $\omega$ to its amplitude \emph{E}, up to a material-dependent constant. In a correlated electron system, a many-body analog of the Keldysh parameter is defined as
\begin{equation}
    \gamma_K = \frac{\hbar\omega}{eE\xi}
\end{equation}
where $\xi$ is the electron-hole correlation length \cite{oka2012nonlinear}. Using the effective Coulomb interaction from density functional theory and the optical gap values \cite{chittari2016electronic,kim2018charge}, we estimate the correlation strength to be $U/\tau\sim7.3$, thus locating the system in the strong correlation regime and justifying the adoption of the above definition of $\gamma_K$. Analytical theory on one-dimensional Hubbard model suggests $\xi=\ln^{-1}{(U/4.377\tau)}a$ \cite{stafford1993scaling}. In the driven Hubbard Hamiltonian, the optical field enters through the Peierls phase, while the bare parameters $U$ and $\tau$ remain fixed in our definition of $\gamma_K$. Thus, $\gamma_K$ here is a fixed constant given the drive parameters. Due to the lack of knowledge about $\xi$ in the Hubbard model on a honeycomb lattice, we use this equation to approximate $\xi=1.96a$ throughout this work. In addition, the dielectric screening of the sample in the sub-gap regime reduces the applied electric field $E_a$ to only half of its vacuum strength inside the sample ($\epsilon_1\approx9$ is used to calculate field penetration \cite{piacentini1982optical,kim2018charge}, which is also supported by density functional calculations \cite{lane2020thickness}). Following the convention from literature, we still use the vacuum peak electric field $E$ to indicate the applied field strength $E_a$ in the main text as well as in Supplementary Materials, but adopt $E_{\rm{int}}=E_a/2$ to calculate the Keldysh parameter $\gamma_K$ and ponderomotive energy $E_p$. We note that the absolute numerical value of $\gamma_K$ should nevertheless be regarded as an estimate because the one-dimensional correlation length is used as a proxy for the corresponding length scale in the multiorbital quasi-two-dimensional NiPS$_3$. This model dependence may slightly shift the numerical placement of the crossover in Keldysh space, but it does not alter the measured field- and photon-energy-dependent signatures used to identify the nonadiabatic crossover.

Strictly speaking, the Keldysh theory is formally applicable to a system driven by a monochromatic continuous-wave laser \cite{keldysh1964ionization, popov2004tunnel} (see discussions in Sec.~\ref{sec:Keldysh theory}). However, in most solid-state experiments, Gaussian laser pulses are often utilized to achieve a high peak field strength. Moreover, because Gaussian beams are the lowest-order self-consistent field distribution in optical resonators provided that there are no intracavity elements causing beam distortions, the output beams of many modern lasers are Gaussian. These two factors naturally lead to a non-uniform and inhomogeneous generation of electron-hole pairs throughout the excitation volume as a function of time and real-space coordinates. For this reason, below we simulate three situations with different laser parameters that correspond to distinct regimes in the Keldysh space and compare the results from a realistic laser pulse with those from a uniform monochromatic continuous-wave laser.

Let us first model the electric field amplitude distribution at the sample as
\begin{equation}\label{eq:s1}
    E(r,t) = E_0\exp(-\frac{t^2}{2\sigma_t^2})\exp(-\frac{r^2}{2\sigma_r^2})
\end{equation}
where $\sigma_t$ and $\sigma_r$ are the standard deviations of the normal distribution in the temporal and spatial domain and will be taken as unity in numerical simulations. The peak Keldysh parameter, which is broadly adopted in literature and current work, is defined as
\begin{equation}\label{eq:s2}
    \gamma_{K,\rm{peak}} = \frac{\hbar\omega}{eE_0a}
\end{equation}
and the instantaneous Keldysh parameter is defined as
\begin{equation}\label{eq:s3}
    \gamma_K(r,t) = \frac{\hbar\omega}{eE(r,t)a}.
\end{equation}
For simplicity, we assume that the instantaneous interband transition only depends on the instantaneous Keldysh parameter $\gamma_K(r,t)$ and ignore any coherent generation of transient quasiparticles. Since narrow-band laser pulses were exploited in experiments, we omit chirping of the pulse which gives rise to a time-dependent frequency $\omega(t)$ and use its central frequency $\omega$ for estimation.

\subsection{$\gamma_{K,\rm{peak}}>1$}
It is clear from Eqns.~\ref{eq:s1}-\ref{eq:s3} that $\gamma_{K,\rm{peak}}\leq\gamma_K(r,t)$. Therefore, $\gamma_K(r,t)>1$ holds true for any variables in this regime, i.e., the multiphoton regime. Suppose the order of the multiphoton process is \emph{a}, we have the electron-hole pair density
\begin{equation}
    n \propto \int rE^{2a}\,dtdr \ =\int rE_0^{2a}\exp(-\frac{at^2}{\sigma_t^2})\exp(-\frac{ar^2}{\sigma_r^2})\,dtdr \ \propto E_0^{2a}
\end{equation}
Hence, a multiphoton process strictly preserves itself in a realistic laser setting.

\subsection{$\gamma_{K,\rm{peak}}\lesssim1$}
In $\gamma_{K,\rm{peak}}\lesssim1$ regime, we show that although \emph{n} is nominally dominated by the tunneling process, there is still significant contribution from the multiphoton channel. As an example, we take $\gamma_{K,\rm{peak}}=0.7$ and $\hbar\omega/\Delta=0.3$, which are close to experiment parameters in Fig.~\ref{fig:m3}a-c.

Figure~\ref{fig:Keldysh parameter}a displays the temporal and spatial distribution of $1/\gamma_K=eEa/\hbar\omega$ in a Gaussian laser pulse with a Gaussian beam profile. We can see that the tunneling process only dominates dynamics near the center of the pulse and the beam and the multiphoton process (in this case, a four-photon process) takes place elsewhere. In this toy model, we assume the following dependencies of \emph{n} on instantaneous envelope field amplitude $E(r,t)$
\begin{equation}\label{eq:s4}
    n\propto
    \begin{cases}
        E^8 &(\gamma_K>1)\\
        E^2\exp{(- E_{\rm{th1}}/E)} &(\gamma_K<1)
    \end{cases}
\end{equation}
and coefficients in the two regimes are chosen such that $n$ is continuous across $\gamma_K=1$ (Fig.~\ref{fig:Keldysh parameter}b). Then, we integrate it over time and area to derive the relation between \emph{n} and $E_0$ in a Gaussian pulse. Qualitatively, the results (Fig.~\ref{fig:Keldysh parameter}c) appear similarly to Eqn.~\ref{eq:s4}: Power-law fitting in the multiphoton regime starts to deviate the data points across the Keldysh crossover line ($\gamma_{K,\rm{peak}}=1$) and the tunneling functional form describes the data better. Nevertheless, the fitted tunneling threshold significantly differs from the input as $E_{\rm{th1,fit}}=2.2E_{\rm{th1}}$. Such a discrepancy can be reconciled if we project \emph{n} onto the multiphoton and tunneling channels, respectively. As shown in Fig.~\ref{fig:Keldysh parameter}d, in the tunneling regime near the Keldysh crossover, photocarriers generated by the multiphoton process even dominate over those by quantum tunneling. At $\gamma_{K,\rm{peak}}=0.7$, multiphoton-triggered electron-hole pairs still contribute to 25\% of the total carriers. We conjecture that this phenomenon contributes to the anisotropic in-plane electron-hole pair density shown in Fig.~\ref{fig:m3}b.

\subsection{$0\ll\gamma_{K,\rm{peak}}<1$}
$0\ll\gamma_{K,\rm{peak}}<1$ is the nonadiabatic tunneling regime as primarily discussed throughout this work. Here, we demonstrate that when $\gamma_{K,\rm{peak}}$ and $\hbar\omega/\Delta$ are substantially smaller than 1, the influence from the Gaussian distribution of the laser pulse is negligible and conclusions drawn from taking the peak electric field $E_0$ as the single parameter to describe the dynamics are reliable.

\begin{figure}[!htb]
\includegraphics{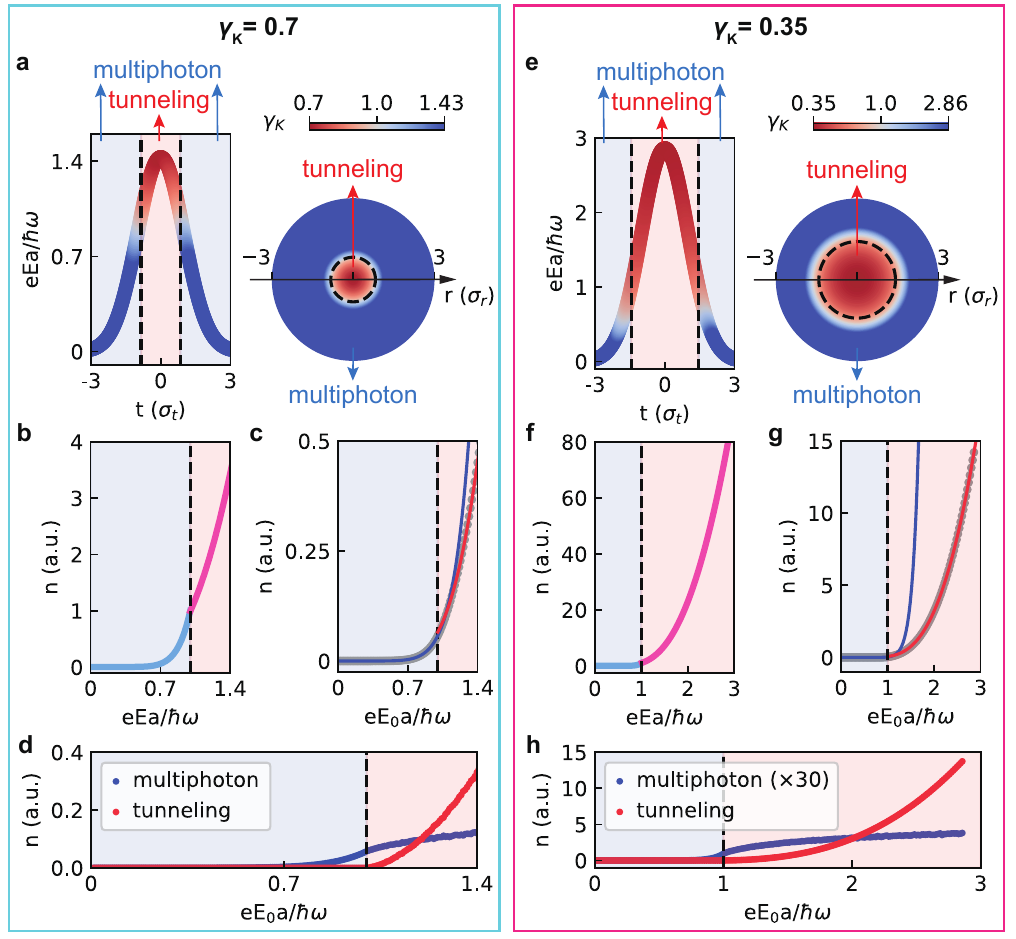}
\caption{\label{fig:Keldysh parameter} Numerical simulations of the electron-hole pair density \emph{n} with different laser parameters. (a) Temporal and spatial distribution of $1/\gamma_K=eEa/\hbar\omega$ in a Gaussian laser pulse with a Gaussian beam profile ($\gamma_{K,\rm{peak}}=0.7$). The left panel shows its time evolution at $r=0$ and the right panel shows its spatial distribution at $t=0$. Blue and red shaded regions correspond to instantaneous multiphoton ($\gamma_K>1$) and tunneling ($\gamma_K<1$) regimes, respectively. (b) \emph{n} as a function of $eEa/\hbar\omega$ in a continuous-wave laser. (c) \emph{n} as a function of $eE_0a/\hbar\omega$ in a Gaussian-pulsed laser with a Gaussian beam profile. Blue and red lines are fit to the data in multiphoton and tunneling regimes, respectively. (d) \emph{n} generated by multiphoton and tunneling processes as a function of $eE_0a/\hbar\omega$ in a Gaussian-pulsed laser with a Gaussian beam profile. (e)-(h) Same as (a)-(d) but with $\gamma_{K,\rm{peak}}=0.35$. Dashed lines in each panel denote the Keldysh crossover.}
\end{figure}

To echo experimental results in Fig.~\ref{fig:m3}d-i, we adopt $\gamma_{K,\rm{peak}}=0.35$ and $\hbar\omega/\Delta=0.18$. Figure~\ref{fig:Keldysh parameter}e again shows the temporal and spatial distribution of $1/\gamma_K=eEa/\hbar\omega$ in a Gaussian laser pulse with a Gaussian beam profile. Although contribution from the multiphoton channel (in this case, a six-photon process) is always present, its high order and low field intensity make its role minor in the entire generation process, as will be elaborated below.

We assume \emph{n} as a function of instantaneous envelope field amplitude \emph{E} at this wavelength described by
\begin{equation}\label{eq:s5}
    n\propto
    \begin{cases}
        E^{12} &(\gamma_K>1)\\
        E^2\exp{(- E_{\rm{th2}}/E)} &(\gamma_K<1)
    \end{cases}
\end{equation}
and the continuity is again ensured by choosing appropriate prefactors. As shown in Fig.~\ref{fig:Keldysh parameter}f, \emph{n} grows rapidly in the tunneling regime. After integration, the result yet again parallels with Eqn.~\ref{eq:s5} (Fig.~\ref{fig:Keldysh parameter}g) while there is still quantitatively considerable mismatch between the fitted and input tunneling threshold: $E_{\rm{th2,fit}}=1.26E_{\rm{th2}}$. We rule out the origin of the disagreement by calculating \emph{n} triggered in distinct channels. Figure~\ref{fig:Keldysh parameter}h indicates that the number of multiphoton-generated carriers is only less than 1\% of the total \emph{n}. Consequently, we attribute this numerical inconsistency in $E_{\rm{th2}}$ to the Gaussian nature of the pulse and the beam profile, which inevitably mixes a broad range of field strengths in photocarrier generation. Nonetheless, due to the overwhelming role of tunneling-generated carriers in the tunneling regime here, it is justified to draw conclusions only based on this process without the need to take a six-photon absorption channel into consideration.

We also briefly comment on the advantage of our transient reflectivity protocol employed to characterize nonlinear absorptions here. Previous approaches to this problem in solids include the z-scan technique \cite{sheik2002sensitive} and photoluminescence \cite{zhang2025keldysh}. However, these methods have several limitations which make them inapplicable here. Firstly, the open-aperture z-scan configuration relies on the assumption of real-space distribution of light intensity propagating through the sample to characterize its nonlinear absorption coefficient. However, the order of a multiphoton process can only be inferred from its relation to the bandgap in the multiphoton regime, while our present work focuses on the nonadiabatic tunneling regime. The strongly non-perturbative nature of light-matter interaction here renders the material response in real space not a simple function of nonlinear refractive indices and absorption coefficients. Secondly, in the z-scan or photoluminescence measurements, all light transmitting through or reflecting from the sample is collected by the detector, while from the numerical simulations above, this will contain contributions from multiple channels near the Keldysh crossover, making the data interpretation less straightforward. Thirdly, the photoluminescence signal from NiPS$_3$ has a broad spectral distribution from 1.3 eV to 1.5 eV \cite{kang2020coherent}. This will strongly mix with signals from high harmonic generation of the pump, which are usually broad-band as well, especially driven by an MIR laser. Since high harmonics may have different scaling from incoherent absorption, their overlap will render the determination of $n$ from bare emission unreliable under current experiment conditions.

In contrast, transient reflectivity measurements are free from any assumptions of the laser beam distribution and interference from high harmonic photons, and the large mismatch between the pump and probe beam sizes adopted in our experimental setting ensures that only signals close to the pump beam center are collected by the probe, making the adoption of peak electric field $E$ as a single parameter to describe the laser field strength more reasonable.

\clearpage
\section{Hubbard model simulations}
\label{sec:Hubbard}
We adopt a single-band Hubbard model at half-filling, on a square lattice, as a prototype of correlated electron systems:
\begin{equation} \label{eq:supplementary_hubbard_hamiltonian}
    H(t) = -\tau\sum_{j, \hat{\delta}, \alpha}  e^{ie \hat{\delta} \cdot \vec{A}(t)} \; c_{j+\hat{\delta}, \alpha}^\dagger c_{j, \alpha} + U \sum_j n_{j,\uparrow} n_{j,\downarrow}.
\end{equation}
Here, $\tau$ is the nearest-neighbor electron hopping amplitude, lattice sites are indexed by $j$, the unit vector $\hat{\delta} \in \{\pm x, \pm y\}$ denotes nearest-neighbor directions, $U$ is the on-site Coulomb repulsion, and $\alpha$ is the electron spin. The pulsed laser field is represented by the time-dependent vector potential
\begin{equation}
    A(t) = A_0 \; e^{-\frac{(t - t_0)^2}{2\sigma^2}}\; \cos{(\omega t)},
\end{equation}
which enters the Hamiltonian~\eqref{eq:supplementary_hubbard_hamiltonian} via Peierls’ substitution. We remark that using a square lattice as a minimal prototype of a gapped correlated electron system for the simulation, rather than the honeycomb lattice of NiPS$_3$, is expected to alter the absolute amplitude, anisotropy, and detailed resonance lineshape of photoexcitation rates, but not the underlying physical mechanisms; indeed, the multiphoton and tunneling mechanisms discussed in this work, as well as their interplay and cooperation, are not reliant on a particular lattice geometry (Sec.~\ref{sec:Keldysh theory}). For the numerical simulation, we employ $U=10 \tau = 1.93$~eV, as well as a pulse intensity duration of $T_{\rm{pump}}=100$~fs taken to match the experiment conditions.

We simulate the system's nonequilibrium dynamics under optical driving, starting from the Mott insulating ground state, by using a time-dependent Gaussian variational approach. In the Mott state at half-filling, charge localization of one electron per site gives rise to insulating behavior, while the local magnetic moments interact antiferromagnetically via the superexchange coupling $J = 4 \tau^2 / U$. This latter interaction gives rise to strong antiferromagnetic (AFM) correlations at short ranges, characterized by the wavevector $\mathbf{Q} = (\pi, \pi)$, and even long-range AFM order in the case of layered materials below the N\'{e}el temperature. In any case, the ultrafast doublon-hole pair dynamics relevant for this work always occurs on very short length scales: for example, the real-space picture of Landau-Zener tunneling involves one electron tunneling to another nearby site (at most a few lattice constants away), gaining the electrostatic energy due to the applied electric field, and in exchange paying the energy cost $U$ to form a doublon. The short-range nature of the doublon-hole dynamics means that it always takes place in a background of AFM spin correlations, and therefore we can assume for simplicity ordering at wavevector $\mathbf Q$ throughout our approach.

The above arguments motivate the consideration of the following time-dependent variational wavefunction:
\begin{equation} \label{eq:supplementary_variational_wavefunction}
    |\psi(t) \rangle = \prod_{\bold{k}\in\rm{BZ^\prime},\alpha} \Big( u_{\bold{k}} (t) \; c_{\bold{k}, \alpha}^\dagger + v_{\bold{k}}(t) \; c_{\bold{k+Q}, \alpha}^\dagger \Big) |0\rangle,
\end{equation}
where $|0\rangle$ is the fermionic vacuum state, and BZ$^\prime$ denotes the magnetic Brillouin zone, which is folded in half with respect to the structural one, due to AFM order effectively doubling the unit cell. We choose the initial conditions $u_{\bold{k}} (0), v_{\bold{k}}(0)$ to minimize the energy expectation $\langle \psi(0) | H | \psi(0) \rangle$, i.e. we begin in the variational ground state. This yields the coefficients 
\begin{equation}
    u_{\bold{k}} (0) = \sin \left( {\pi \over 4} + {\varphi_\mathbf{k} \over 2} \right), \qquad \qquad v_{\bold{k}} (0) = \sin \left( {\pi \over 4} - {\varphi_\mathbf{k} \over 2} \right),
\end{equation}
which represent a Bogoliubov rotation by the angle $\varphi_\mathbf{k} \equiv \arctan [2 \epsilon_{\mathbf k} / U S]$. Here, $\epsilon_{\bold{k}} = -2 \tau (\cos{k_x} + \cos{k_y})$ is the kinetic energy on a square lattice with nearest-neighbor hopping, and $S=\langle\psi|S_{\bold{Q}}^z|\psi\rangle$ is the N\'{e}el order parameter given by the expectation of the operator
\begin{equation}
    S_{\bold{Q}}^z = \frac{1}{N} \sum_{\bold{k}\alpha}\alpha \; c_{\bold{k+Q}, \alpha}^\dagger c_{\bold{k}, \alpha}.
\end{equation}
The order parameter in equilibrium obeys the self-consistency equation
\begin{equation}
    \frac{1}{N}\sum_{\rm{BZ^\prime}}\frac{1}{\sqrt{(\epsilon_{\bold{k}}/U)^2 + (S/2)^2}} = 1,
\end{equation}
and it also contributes to setting the quasiparticle dispersion on top of the variational ground state,
\begin{equation} \label{eq:supplementary_quasiparticle_dispersion}
    E_\mathbf{k} = \sqrt{\epsilon_{\mathbf k}^2 + ({US / 2})^2}.
\end{equation}
The doublon-hole pair density of states, at zero total momentum, is shown in Fig.~\ref{fig:density_of_states} below. The sharp peak at the bottom of the band is an artifact of restricting hopping to nearest neighbors on a square lattice. As discussed in the main text, this makes the simulated multiphoton thresholds appear as peaks rather than steps in plots of carrier density versus frequency.
\begin{figure}[!h]
	\centering	
    \scalebox{0.5}{\includegraphics{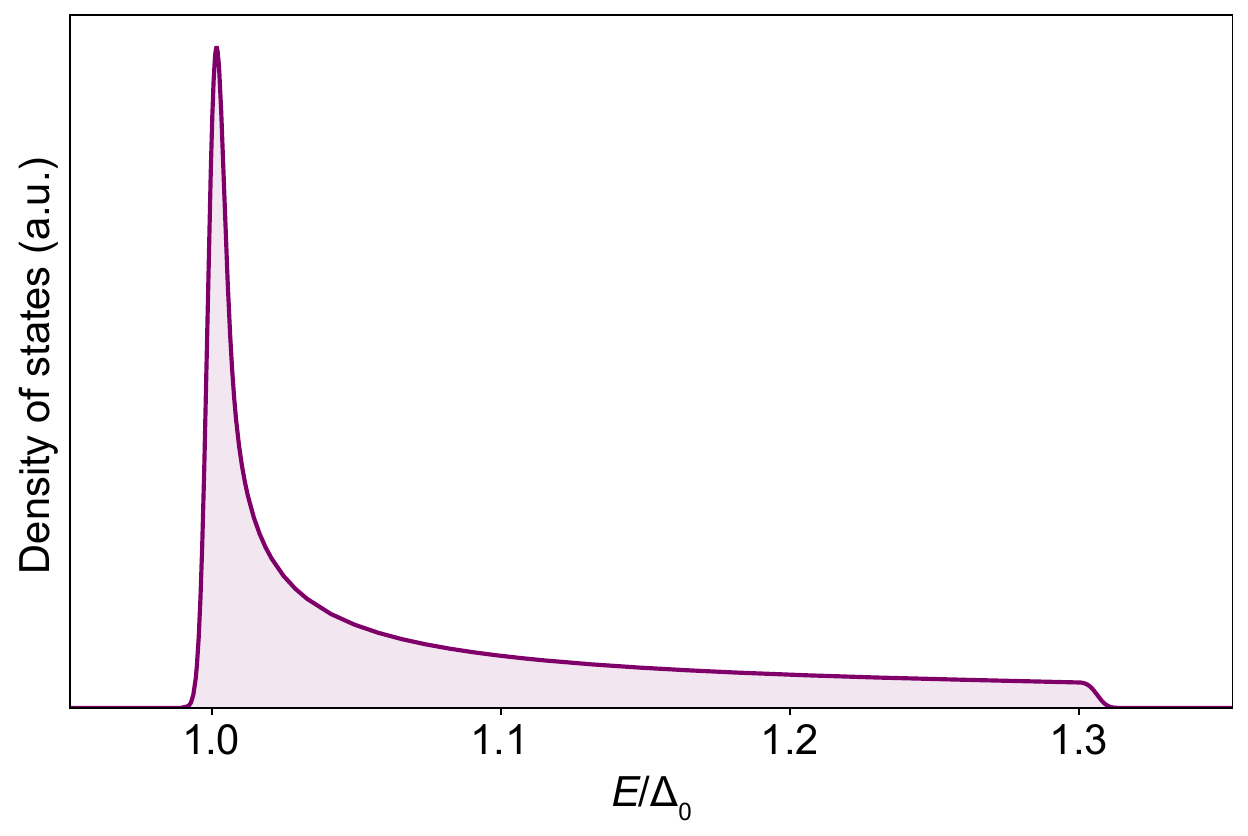}}
	\caption{Density of states in the Hubbard model. Typical shape for the density of states, for doublon-hole pair excitations, at zero total momentum; calculated for $U / \tau \approx 10$. The sharp peak at the bottom of the band is a consequence of assuming hopping to be nearest-neighbor only. This figure is adapted from \cite{andrei2025subgap}.}
	\label{fig:density_of_states}
\end{figure}

Once the variational ground state is found, one may write evolution equations for the coefficients $u_{\bold{k}} (t)$ and $v_{\bold{k}} (t)$ under optical driving. On the other hand, formulating the time evolution in terms of a pure wavefunction implicitly ignores decoherence effects intrinsic to all solid-state systems. In order to include these as well, we reformulate the ansatz \eqref{eq:supplementary_variational_wavefunction} in terms of a time-dependent density matrix. Also note that, upon turning on the drive $A(t)$, the effective kinetic energy is modified to $\epsilon_{\bold{k}}(t) = -2 \tau \left[ \cos{(k_x -e A_x(t) \; a)} + \cos{(k_y -e A_y(t) \; a)} \right]$. In consequence, we also promote $S(t)$, $\varphi_\mathbf{k}(t)$, $E_\mathbf{k}(t)$ etc. to time-dependent functions. Applying the Bogoliubov transformation in order to work in the doublon-hole basis directly, we obtain the density-matrix ansatz
\begin{equation}
    \rho_\text{var}(t) =\bigotimes_{\substack{\mathbf k \in \rm{BZ}^\prime, \alpha}} \begin{pmatrix} {1 \over 2} [{1 - \rho_{\mathbf k}^z (t)}] & \alpha \rho_{\mathbf k}^+ (t) \\ \alpha \rho_{\mathbf k}^- (t) & {1 \over 2} [{1 + \rho_{\mathbf k}^z (t)}] \end{pmatrix},
\end{equation}
with initial conditions $\rho_{\mathbf k}^z (0) = 1$ and $\rho_{\mathbf k}^+ (0) = \rho_{\mathbf k}^- (0) = 0$. Note that $[\rho_{\mathbf k}^+]^* = \rho_{\mathbf k}^-$ holds at all times. The coupled evolution equations for the above coefficients, as well as the order parameter $S$, are
\begin{subequations} \label{eq:supplementary_evolution_equations}
\begin{align}
	\partial_t \rho_\mathbf{k}^+ &= {\partial_t \varphi_\mathbf{k} \over 2} \; \rho_\mathbf{k}^z + 2 [i E_\mathbf{k} - \Gamma]  \rho_\mathbf{k}^+, \\ 
	\partial_t \rho_\mathbf{k}^z &= - (\partial_t \varphi_\mathbf{k}) \; (\rho_\mathbf{k}^+ + \rho_\mathbf{k}^-), \\ 
    \partial_t S &= -{4 \over N_{\rm{BZ}^\prime}} \sum_{\mathbf k \in {\rm{BZ}}^\prime} \epsilon_{\mathbf k} \left[ \rho_{\mathbf k}^{y} +  {\Gamma \over 2 E_{\mathbf k}} \rho_{\mathbf k}^x \right].
\end{align}\label{eq:evolution eqns}
\end{subequations}
Here, $\Gamma = 1.64$ meV represents a small decoherence rate, significantly below all other energy scales in the system. We mention that, within this order of magnitude, the precise value of $\Gamma$ is irrelevant for the photoexcitation regimes considered in this work; however, the $\Gamma \to 0^+$ limit is difficult to reach numerically, so we employ the small value mentioned above. The observable of interest is the doublon-hole pair density,
\begin{equation}
    n_{\rm{qp}}(t) = \frac{1}{N_{\rm{BZ}^\prime}}\sum_{\bold{k}\in{\rm{BZ}}^\prime}\frac{1-\rho_{\bold{k}}^z(t)}{2}\label{eq:nqp},
\end{equation}
at times $t > T_{\rm pump}$ right after the pump excitation. While the full numerical solution of Eqns.~\eqref{eq:supplementary_evolution_equations} is presented in the main text, we remark that an approximate analytic treatment in the limit $\Gamma \to 0^+$ allows us to recover both the multiphoton and tunneling photoexcitation rates,
\begin{subequations}
    \begin{align}
        n_\text{multiphoton} &\propto \left( {e E_0 a \over \hbar \omega} \right)^{2 \Delta / \omega} \; \rho(n_\text{ph} \; \omega), \label{eq:nqpa}\\
        n_\text{tunnel} &\propto E_0^2 \; \exp \left[ - \gamma(\omega) \; {\Delta \over e E_0 a} \right].\label{eq:nqpb}
    \end{align}
\end{subequations}
Here $\Delta$ is the Mott gap, $E_0$ is the pump field, $\omega$ is the driving frequency, $n_{\rm ph} = \lceil {\Delta \over \omega} \rceil$ is the order of the multiphoton process, and $\gamma(\omega)$ is a frequency-dependent tunneling rate, showing significant enhancement around multiphoton thresholds. The expressions above therefore capture the important qualitative features of the experiment, discussed in the main text. 

Moreover, Eqns.~\ref{eq:nqpa} and~\ref{eq:nqpb} also make the separation between lattice-specific and generic ingredients explicit. In the multiphoton contribution, the doublon-holon density of states $\rho(n_\text{ph} \; \omega)$ directly controls the resonance weight, while the optical matrix elements, which are not displayed in the simplified expression, additionally determine its polarization dependence. The square-lattice band-edge singularity therefore makes the calculated photon-order structure unusually sharp and concentrates their spectral weight near the gap edge. Switching to a smoother dispersion on the honeycomb lattice would broaden and redistribute these structures. By contrast, the exponential field dependence of the tunneling contribution is a generic strong-field form, although $\gamma(\omega)$ and its prefactors retain quantitative dependence on the exact dispersion and interaction parameters. The crossover to non-perturbative field dependence and the persistence of multiphoton-like structure near this crossover consequently do not require the square-lattice density-of-states singularity, although their quantitative amplitudes and anisotropies remain lattice-dependent.

The calculated redshift in the spectrum (Fig.~\ref{fig:m4}a) has a distinct, dynamical origin. In the time-dependent variational equations, the drive produces transient doublon-holon occupation even when its photon energy is below the static threshold. This photoexcitation changes the AFM order parameter $S(t)$ and hence the self-consistent correlation gap $\Delta(t)=US(t)$. The reduction of $S(t)$ lowers the instantaneous gap, so that a pump initially below a static multiphoton threshold can be brought into resonance during the pulse. The static density-of-states singularity enhances and sharpens the resulting response, and may affect how the position of its maximum is identified, but it does not itself generate the time-dependent reduction of $\Delta(t)$. A smoother dispersion would therefore change the width and spectral weight of the calculated resonance and could alter the relative visibility of this redshift and the ponderomotive blueshift, without removing the self-consistent gap-feedback mechanism.

For more details about the variational approach, numerical simulations, and analytic treatment, we refer the readers to \cite{andrei2025subgap}.

\clearpage
\section{Keldysh paradigmatic numerical simulations}
\label{sec:Keldysh theory}
Here we show that our key experimental signatures can also be qualitatively captured by a toy model simulation within the Keldysh framework \cite{keldysh1964ionization}, which is briefly reviewed below.

Before introducing the excitation rate, we summarize the approximations used by Keldysh to obtain the closed analytic expression.

\noindent\textit{(i) Monochromatic, homogeneous drive.}
The optical field is treated as classical, spatially uniform, linearly polarized, and monochromatic, $F(t)=F\cos(\omega t)$. The transition rate is evaluated in the long-time limit and averaged over many field cycles. Finite pulse duration, bandwidth, and propagation are therefore absent in this formalism.

\noindent\textit{(ii) Fixed two-band strong-field approximation.}
Excitation is represented as a direct vertical transition from one fixed valence band to one fixed conduction band. Intraband acceleration is included nonperturbatively through the accelerated momentum $\mathbf p(t)=\mathbf p+(e\mathbf F/\omega)\sin\omega t$ whereas interband transfer is evaluated through the leading transition matrix element. Carrier depletion, Pauli blocking, scattering, dephasing, self-consistent gap dynamics, and additional bands are not included. Electron-hole attraction and intermediate excitonic or defect channels are also excluded from the direct excitation rate.

\noindent\textit{(iii) Leading complex-saddle contribution.}
The contour integral for the transition amplitude is evaluated by steepest descent at complex points satisfying
\begin{equation}
\varepsilon\!\left(\mathbf p+\frac{e\mathbf F}{\omega}u_s\right)=0.
\label{eq:r27_saddle}
\end{equation}
The leading complex action and fluctuation prefactor are retained, whereas higher-order saddle corrections and nonsaddle contour contributions are neglected. In the corresponding derivation, Keldysh explicitly assumes \emph{$\hbar\omega/\Delta$ to be sufficiently small}. The closed saddle-evaluated expression is therefore quantitatively controlled in the strongly subgap limit $\hbar\omega/\Delta\ll1$.

\noindent\textit{(iv) Singular matrix element and Kane dispersion.}
Near the complex band crossing, the interband matrix element is replaced by its singular pole contribution. A closed expression is then obtained by adopting the isotropic two-band Kane dispersion \cite{kane1957band}
\begin{equation}
\varepsilon(\mathbf p)=\Delta\sqrt{1+\frac{p^2}{m\Delta}},
\label{eq:r27_kane}
\end{equation}
with a fixed direct gap $\Delta$ and reduced mass $m$. The resulting formula therefore describes this idealized band model rather than an arbitrary material dispersion.

\noindent\textit{(v) Small canonical-momentum expansion.}
The canonical momentum $\mathbf p$ contributing to the final integral is assumed to be small, although the field-driven displacement $\mathbf p+e\mathbf F u/\omega$ is not. The cycle-averaged energy and complex action are expanded through quadratic order in $p_{\parallel}$ and $p_{\perp}$, after which the momentum integral is evaluated in the continuum approximation.

\noindent\textit{(vi) Averaging of saddle-point interference.}
The term linear in $p_{\parallel}$ has opposite signs for the two relevant saddle points. Keldysh replaced the resulting rapidly oscillating interference factor by its momentum average, thereby removing the detailed interference between the two saddle contributions. Retaining this interference has subsequently been shown to restore odd-even multiphoton selection rules and improve the quantitative description of low-order absorption~\cite{shcheblanov2017nonlinear}.

With the definition of the Keldysh parameter $\gamma$ in Eq.~\ref{eq:gk}, Keldysh gave the expression of excitation probability as
\begin{equation}
    w = \frac{2\omega}{9\pi}(\frac{\sqrt{1+\gamma^2}}{\gamma}\frac{m\omega}{\hbar})^{3/2}Q(\gamma, \frac{\widetilde{\Delta}}{\hbar\omega})\exp\{-\pi\langle\frac{\widetilde{\Delta}}{\hbar\omega}+1\rangle[K(\frac{\gamma}{\sqrt{1+\gamma^2}})-E(\frac{\gamma}{\sqrt{1+\gamma^2}})]/E(\frac{1}{\sqrt{1+\gamma^2}})\}\label{eq:Keldysh}
\end{equation}
where $\langle n\rangle$ denotes the integer part of $n$ and $\widetilde{\Delta}$ is the laser-dressed band gap
\begin{equation}\label{eq:s6}
    \widetilde{\Delta} = \frac{2}{\pi}\Delta\frac{\sqrt{1+\gamma^2}}{\gamma}E(\frac{1}{\sqrt{1+\gamma^2}}).
\end{equation}
The functions $K$ and $E$ are complete elliptic integrals of the first and second kind. The function $Q$ encodes the spectrum
structure from the discreteness of the number of absorbed photons and is defined as
\begin{align}
    Q(\gamma,x) = [\pi/2K(\frac{1}{\sqrt{1+\gamma^2}})]^{1/2}\sum_{n=0}^{\infty}\exp\{-n\pi[K(\frac{\gamma}{\sqrt{1+\gamma^2}})-E(\frac{\gamma}{\sqrt{1+\gamma^2}})]/E(\frac{1}{\sqrt{1+\gamma^2}})\}\\
    \Phi\{[\pi^2(2\langle x+1\rangle-2x+n)/2K(\frac{1}{\sqrt{1+\gamma^2}})E(\frac{1}{\sqrt{1+\gamma^2}})]^{1/2}\}
\end{align}
where the function $\Phi$ is defined as
\begin{equation}
    \Phi(z) = \int_0^z e^{y^2-z^2}dy.
\end{equation}
We note that none of the approximations above takes the quasi-static limit, although the saddle-point evaluation of the general expression Eqn.~\ref{eq:Keldysh} relies on the strong subgap approximation. This requirement depends only on frequency and is distinct from the adiabaticity condition $\gamma_K\ll1$. We therefore use the calculation as a paradigmatic finite-frequency description of the crossover and its field-dressed resonances, rather than as a parameter-free prediction of the absolute excitation rate in NiPS$_3$.

In addition, the expressions above make an important distinction between the material parameters defining the Keldysh parameter and the field-dressed threshold entering the excitation probability. Within the Kane dispersion (Eqn.~\ref{eq:r27_kane}) adopted by Keldysh, Eqn.~\ref{eq:gk} is defined using the undriven reduced electron-hole mass $m$ and bandgap $\Delta$. These parameters remain fixed in the definition of $\gamma_K$. On the other hand, the laser-dressed threshold $\widetilde{\Delta}$ (bandgap renormalization) in Eqn.~\ref{eq:s6} is a separate quantity. In the multiphoton limit, it approaches
\begin{equation}
\widetilde{\Delta}
=\Delta+\frac{e^2E^2}{4m\omega^2}
+\mathcal{O}\!\left(\Delta\gamma_K^{-4}\right)
=\Delta+E_p+\mathcal{O}\!\left(\Delta\gamma_K^{-4}\right).
\label{eq:dressed_threshold_limit_addition}
\end{equation}
The dressed threshold enters the discrete channel factor $Q$ and the excitation probability $w$, while $\gamma_K$ continues to be evaluated using the undriven $m$ and $\Delta$. This separation permits the photon-order resonances to shift with field while the Keldysh parameter continues to organize the crossover between multiphoton- and tunneling-like excitation.

\begin{figure}[!htb]
\includegraphics{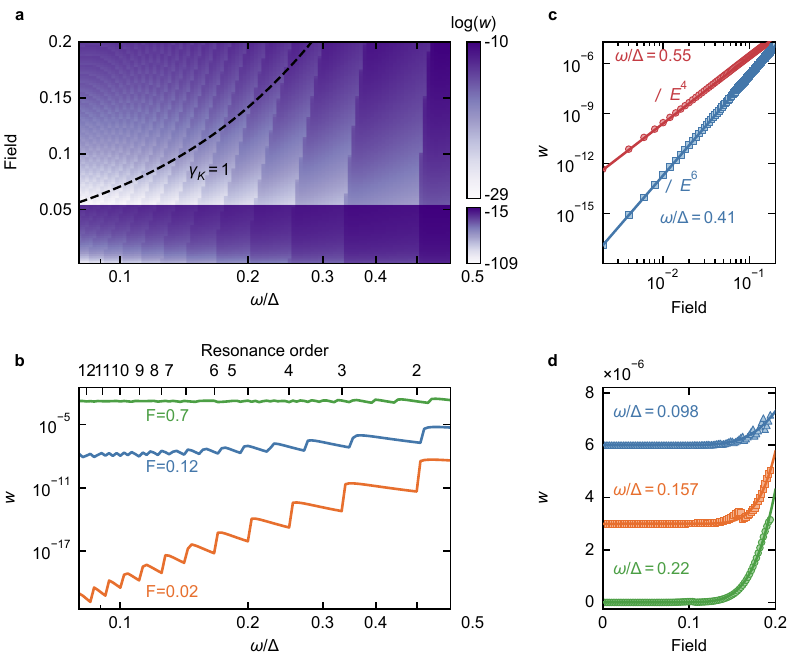}
\caption{\label{fig:Keldysh theory} Numerical simulations of the excitation probability $w$ with different laser parameters. (a) Simulated $w$ as a function of normalized light frequency $\omega/\Delta$ and field amplitude $F$ on a log scale. The black dashed line denotes $\gamma_K=1$ separating the Keldysh space into multiphoton-like and tunneling-like regimes. The lower and upper part of the color plot are on different scales to highlight the existence of multiphoton-like resonances in the multiphoton and nonadiabatic tunneling regimes. (b) Selective constant electric field cuts at $F = 0.02, 0.12, 0.7$, corresponding to multiphoton, multiphoton-to-nonadiabatic tunneling, and adiabatic tunneling regimes, respectively. (c) Selective constant photon energy cuts at $\hbar\omega/\Delta = 0.55, 0.41$ overlaid by their power-law fitting on a log-log scale, corresponding to two-photon and three-photon processes, respectively. (d) Selective constant photon energy cuts at $\hbar\omega/\Delta = 0.22, 0.157, 0.098$ overlaid by their fitting to the tunneling formula. The curves are vertically offset for clarity.}
\end{figure}

In our numerical simulations, we adopted atomic units for simplicity. Figure~\ref{fig:Keldysh theory}a shows the results of excitation probability $w$ as a function of normalized light frequency $\omega/\Delta$ and field amplitude $F$. Similar to the results obtained with Hubbard model simulations (Fig.~\ref{fig:m4}), multiphoton-like resonances persist well into the tunneling-like regimes without any noticeable discontinuity across the Keldysh crossover (black dashed line in Fig. \ref{fig:Keldysh theory}a). These resonances all exhibit blueshifts in the strong-field regime, originated from the field-induced bandgap renormalization effect (Eqn.~\ref{eq:s6}), which also qualitatively reproduces our experimental observation (Fig.~\ref{fig:m3}). The multiphoton nature of excitations in the $\gamma>1$ regime is verified by power-law fitting the constant energy cuts at $\omega/\Delta=0.55$ and $0.41$, corresponding to two-photon and three-photon processes, respectively (Fig.~\ref{fig:Keldysh theory}c). On the other hand, the electric-field-dependent data in the nonadiabatic tunneling regime could be described by the tunneling functional form well (Fig.~\ref{fig:Keldysh theory}d), signaling their distinct nature. The drop in $w$ at certain field strengths comes from the multiphoton channel closing effect and does not affect our conclusions \cite{kruchinin2018colloquium,story1994channel}.

Figure~\ref{fig:Keldysh theory}b shows constant-field cuts at $F=0.02$, $0.12$, and $0.7$, illustrating the continuous evolution from multiphoton excitation through the finite-frequency nonadiabatic crossover toward the quasi-static tunneling limit. At $F=0.02$, the excitation probability decreases rapidly with increasing photon order and displays well-resolved multiphoton resonances. At $F=0.12$, the low-frequency portion lies on the nominal tunneling side of the crossover, reaching $\gamma_K=0.47$ at the lowest displayed frequency. Its smooth envelope approaches a quasi-plateau, but photon-order structures remain because the calculation is still at finite $\gamma_K$. The much stronger field $F=0.7$ produces a substantially flatter low-frequency response and is included only as a benchmark for the expected approach to the quasi-static limit as it lies far outside the experimental field range. Figure~\ref{fig:Keldysh theory}b therefore demonstrates that frequency dependence is gradually suppressed with decreasing $\gamma_K$, rather than disappearing abruptly at $\gamma_K=1$.

For the two- and three-photon conditions in Fig.~\ref{fig:m2}, $\hbar\omega/\Delta=0.538$ and $0.394$, respectively. We therefore do not use the closed Keldysh expression to predict their absolute excitation rates or resonance weights. Their experimental assignments instead follow directly from the measured $E^4$ and $E^6$ field dependences, the step in the response across $\Delta/2$, and the evolution of the effective fitted exponent from values near $\alpha=3$ below the threshold to values near $\alpha=2$ above it. The low-order Keldysh cuts are retained only as a qualitative consistency check of the photon-channel structure and leading $E^{2N}$ behavior.

The Hubbard and Keldysh calculations accordingly serve complementary purposes. The Hubbard calculation includes self-consistent many-body feedback and demonstrates the coexistence of photon-order structure and tunneling-like field dependence in a correlated insulator. Accordingly, we use its simplified square-lattice dispersion to establish qualitative excitation mechanisms, while treating the direction and magnitude of the experimental resonance shift as material-specific. The Keldysh calculation here does not contain the square-lattice dispersion or its density-of-states singularity, yet it reproduces multiphoton-like resonances across the Keldysh crossover, their tunneling-like field dependence, and the ponderomotive blueshift observed experimentally. The agreement of these complementary approaches on the central crossover signatures shows that our observed phenomena are broadly present in a range of gapped systems.

\clearpage
\section{Possible role of avalanche effects}
\label{sec:impact ionization}
Impact ionization and avalanche multiplication are secondary processes that can, in principle, follow primary strong-field injection~\cite{balling2013femtosecond}. Their physical criterion is distinct from that of direct excitation. The Keldysh parameter compares the characteristic time scale for primary excitation across the gap with the period of the driving field. Avalanche formation begins only after seed carriers have been injected and depends on whether they can acquire sufficient excess energy, undergo an impact event before losing that energy, and repeat this sequence within the pulse. Its efficiency is therefore governed by carrier heating, impact and energy-loss scattering rates, seed density, and pulse duration. Consequently, neither the multi-cycle character of a pulse nor $\gamma_K<1$ by itself establishes appreciable avalanche multiplication.

The frequently used DC energy-gain condition assumes monotonic carrier acceleration and does not determine the multiplication produced by a finite oscillating pulse without the relevant scattering kinetics. Because a material-specific impact coefficient and nonequilibrium scattering rates are not available for NiPS$_3$, we do not attempt to assign an absolute microscopic upper bound to the multiplication factor. Instead, we test whether the measured energy, field, and pulse duration scales are favorable for impact ionization.

In a simple parabolic-band model of impact ionization~\cite{jena2022quantum}, an energetic conduction-band electron or valence-band hole creates an additional electron-hole pair. Energy and momentum conservation give the minimum kinetic energies required for the two processes as
\begin{equation}
\begin{aligned}
E_{\mathrm{II}}^{(e)}
&=
\frac{2m_c^*+m_v^*}{m_c^*+m_v^*}\Delta,
&
E_{\mathrm{II}}^{(h)}
&=
\frac{m_c^*+2m_v^*}{m_c^*+m_v^*}\Delta,
\end{aligned}
\label{eq:impact_kinematic_thresholds_revised}
\end{equation}
where $m_c^*$ and $m_v^*$ are the magnitudes of the conduction- and valence-band effective masses. Using the representative bulk $b$-axis band-curvature masses $m_c^*=3.94m_e$ and $m_v^*=0.51m_e$ for NiPS$_3$~\cite{lane2020thickness}, together with $\Delta=1.8$~eV, gives $E_{\mathrm{II}}^{(e)}=3.39$~eV and $E_{\mathrm{II}}^{(h)}=2.01$~eV.

To determine whether these thresholds can be reached through field acceleration, we consider a carrier injected near a field maximum, as is relevant for tunneling excitation. During the ensuing collisionless quiver cycle, its maximum additional kinetic energy is
\begin{equation}
K_{\max,j}
=2E_{p,j}
=\frac{e^2E_{\mathrm{int}}^2}{2m_j^*\omega^2},
\qquad j=e,h.
\label{eq:maximum_quiver_energy}
\end{equation}
Table~\ref{tab:impact_ballistic_comparison} compares these maximal field-induced kinetic energies with the corresponding impact ionization thresholds for all three experimental excitation conditions (Fig.~\ref{fig:m3}c, f, i).

\begin{table}[h]
\centering
\caption{\textbf{Comparison of the maximal ballistic quiver energies with the impact-ionization thresholds.}
Here, $E_a$ and $E_{\mathrm{int}}$ are the incident vacuum and screened internal peak fields, respectively. The three rows correspond to the field-dependent measurements in Fig.~\ref{fig:m3}c, f, i.}
\label{tab:impact_ballistic_comparison}
\begingroup
\setlength{\tabcolsep}{4pt}
\small
{\begin{tabular}{@{}ccccccc@{}}
\toprule
$\hbar\omega$
& $E_a$
& $E_{\mathrm{int}}$
& $K_{\max,e}$
& $K_{\max,e}/E_{\mathrm{II}}^{(e)}$
& $K_{\max,h}$
& $K_{\max,h}/E_{\mathrm{II}}^{(h)}$ \\
(eV)
& (MV/cm)
& (MV/cm)
& (eV)
& (\%)
& (eV)
& (\%) \\
\midrule
$0.56$ & $14$   & $7.0$ & $0.015$ & $0.44$ & $0.117$ & $5.8$  \\
$0.42$ & $17.8$ & $8.9$ & $0.043$ & $1.27$ & $0.335$ & $16.7$ \\
$0.38$ & $19$   & $9.5$ & $0.060$ & $1.77$ & $0.468$ & $23.3$ \\
\bottomrule
\end{tabular}
}
\endgroup
\end{table}

Throughout these measurements, $K_{\max,e}$ is only $0.015$--$0.060$~eV, corresponding to $0.44\%$--$1.77\%$ of the electron-initiated threshold. The lighter hole provides the more favorable channel, but $K_{\max,h}$ remains between $0.117$ and $0.468$~eV, or $5.8\%$--$23.3\%$ of the hole-initiated threshold. Even under the strongest condition in the table, the maximal hole kinetic energy remains approximately a factor of five below threshold. These estimates therefore rule out prompt, single-cycle ballistic impact ionization as the origin of the measured response. A secondary avalanche contribution would instead require scattering-assisted energy accumulation and carrier multiplication over multiple optical cycles.

We therefore next consider whether such multiplication could develop within the finite pulse duration. Although avalanche breakdown can occur at fields of order MV/cm under DC driving, shortening the drive to sub-picosecond duration leaves much less time for carrier heating and repeated multiplication, thereby significantly raising the required peak field and intensity \cite{rethfeld2017modelling}. Calculations and measurements for insulators with comparable $\Delta/\hbar\omega$ to our driving conditions for NiPS$_3$ indicate onset intensities above approximately $6$~TW/cm$^2$ for pulses of ${\sim}200$~fs~\cite{rajeev2009field}. The maximum vacuum peak intensity in our experiment is only $0.48$~TW/cm$^2$, approximately a factor of $12.5$ lower. Within the dielectric rate-equation models underlying this comparison, such an intensity separation corresponds to an impact-ionization-generated population below $1\%$ of the total photoexcited carrier density. This estimate provides a more realistic finite-pulse consistency check and reinforces the conclusion that avalanche multiplication is minimal here.

Finally, the photon-energy-resolved response helps distinguish primary injection from secondary multiplication. Impact ionization is controlled by whether an already excited carrier acquires sufficient kinetic energy---through its initial excess energy, ponderomotive acceleration, and subsequent scattering dynamics---to create an additional electron-hole pair. Although its efficiency can vary with photon energy through these quantities, impact ionization itself does not impose the discrete resonance condition $N\hbar\omega=\widetilde{E}_g$. It is therefore not expected to generate a sequence of thresholds at integer fractions of the field-renormalized gap. In addition, the multiphoton-like resonance structure and nonperturbative crossover are reproduced by our complementary Keldysh calculation in Sec.~\ref{sec:Keldysh theory}, which contains no impact ionization collision channel, demonstrating that avalanche multiplication is not required to explain either observation.

Taken together, the unfavorable immediate impact energy balance, the one-order-of-magnitude lower intensity than the documented ultrafast multiplication conditions, and the reproduction of the measured signatures without impact collisions indicate that secondary carrier multiplication is minimal under our conditions.

\clearpage

\printsupplementbibliography
\clearpage
\end{bibunit}

\end{document}